\documentclass[twocolumn,showpacs,nofootinbib,aps,prd]{revtex4}

\usepackage{graphicx,url,color}
\usepackage{amsmath,amsthm,amssymb}
\usepackage{multirow}
\usepackage{fmtcount} 
\usepackage{multienum}
\usepackage{gensymb}

\newcommand{\beq}{\begin{equation}}
\newcommand{\eeq}{\end{equation}}

\newcommand{\D}{{\rm d}}
\newcommand{\unit}[1]{\ensuremath{\, \mathrm{#1}}}

\def\lsim{\raise0.3ex\hbox{$\;<$\kern-0.75em\raise-1.1ex\hbox{$\sim\;$}}}
\def\gsim{\raise0.3ex\hbox{$\;>$\kern-0.75em\raise-1.1ex\hbox{$\sim\;$}}}

\providecommand{\e}[1]{\ensuremath{\times 10^{#1}}}
\begin{document}

\title{An Alternative Formation Model for Antideuterons from Dark Matter}

\author{L.~A.~Dal, A.~R.~Raklev$^1$}

\affiliation{$^1$Department of Physics, University of Oslo, Norway}

\begin{abstract}
Antideuterons are a potential messenger for dark matter annihilation or decay in our own galaxy, with very low backgrounds expected from astrophysical processes. The standard coalescence model of antideuteron formation, while simple to implement, is shown to be under considerable strain by recent data from the LHC. We suggest a new empirically based model, with only one free parameter, which is better able to cope with these data, and we explore the consequences of the model for dark matter searches.
\end{abstract}


\maketitle

\section{Introduction}
The use of antideuterons for indirect detection of dark matter (DM) was first suggested in~\cite{Donato:1999gy}. 
Despite the low yield per annihilation or decay, antideuterons can be an important discovery channel due to the extremely low astrophysical background. 
Presently, the AMS-02 experiment is taking data that can improve on current upper bounds for the antideuteron flux at the Earth.

There are several uncertainties at play when calculating the resulting bounds on dark matter models. 
The most significant is the uncertainty in propagation models, while the second is the antideuteron formation model. 
The dark matter halo uncertainty can also be large. We will here concern ourselves with the formation model. 

The formation of antideuterons is commonly described using the so-called coalescence model, which harks back to the 1960s~\cite{Schwarzschild:1963zz,Kapusta:1980zz}. 
In this simple phenomenological model, any antiproton--antineutron pair with momentum difference $|\vec p_{\bar p} - \vec p_{\bar n}| < p_0$, {\it will} combine to form an antideuteron. 
The coalescence momentum $p_0$, typically evaluated in the centre-of-mass (COM) frame of the antinucleons, is a free parameter that must be fixed by calibration against experimental data.
While modified slightly over the years, the coalescence model is still state-of-the-art. 
In calibrating $p_0$ on (relatively) modern experimental data, it has been found that while all available datasets can individually be consistently described by some value of $p_0$, there is no consistent $p_0$ between datasets~\cite{Ibarra:2012cc,Dal:2014nda}. This has been explained by differences in the event generators used to simulate the data, and by the different physical properties of the processes measured, {\it e.g.}\ production from a colorless $e^+e^-$ initial state versus production in $pp$-scattering.

In this work, we will show that new data on deuteron and antideuteron production from the ALICE experiment at the Large Hadron Collider (LHC)~\cite{Serradilla:2013yda}, cannot be well described by the coalescence model. 
Thus, for the first time, challenging the model in a single experiment. 
We will present a new, empirically based model that describes (anti)deuteron formation as a probabilistic process, and show that it is capable of successfully describing the new ALICE data. 
This model will then be applied to make predictions on the antideuteron flux from a generic annihilating dark matter model, and we will compare it to the predictions of the coalescence model. 
Along the way we will also comment on the potential usefulness of future data on deuteron production in order to explore our model further. 

In Section~\ref{sec:Coalescence} we will begin by reviewing the coalescence model and some of its recent modifications. 
We then go on to describe the basis of our new model in Section~\ref{sec:Empirical}. 
In Section~\ref{sec:Calibration}, we proceed by comparing the calibration of the two models on a selection of the available datasets. Section~\ref{sec:DM} describes the resulting cosmic ray antideuteron flux from dark matter in our model, comparing it to the coalescence model, before we conclude in Section~\ref{sec:conclusions}.

\section{The coalescence model} 
\label{sec:Coalescence}

In its initial form, as it was first applied to deuteron production in heavy ion collisions, an additional assumption of isotropic and uncorrelated antiproton and antineutron spectra was used in the coalescence model to obtain an analytical expression for the antideuteron spectrum in terms of the antiproton and antineutron spectra.
These assumptions have, however, been show not to hold in processes relevant to indirect DM detection~\cite{Kadastik:2009ts}, and the coalescence condition should therefore be applied to $\bar p \bar n$-pairs on a per-event basis.

As has been show in Refs~\cite{Ibarra:2012cc,Dal:2014nda}, tuning $p_0$ against experiments measuring different collision processes at differing energy scales does not give a consistent best fit value.
Moreover,  the coalescence model is sensitive to two-particle correlations for (anti)baryons, arising from the structure of the hadronization models in the Monte Carlo event generators used~\cite{Dal:2012my}.
Tuning different event generators to the same experimental data will therefore typically not give the same best fit values for the coalescence momentum.
As discussed in Ref.~\cite{Dal:2014nda}, hadronization parameters in Monte Carlos are usually not tuned to measured two-particle correlations, where they exist, nor indeed with any specific emphasis on reproducing (anti)nucleon spectra.
Tuning hadronization parameters specifically towards antideuteron production is therefore a prospective way of achieving better consistency in fits to experimental data, as well as better agreement between different Monte Carlos.

It was pointed out by the authors of Ref.~\cite{Ibarra:2012cc}, that spatial separation should also be taken into account when evaluating the coalescence condition.
Nuclear interactions take place on scales of a few femtometers, while weakly decaying particles will typically have macroscopic decay lengths.
Their decay products will therefore be produced too far from the primary vertex to have a chance of interacting with particles produced at the primary vertex.
For this reason, weakly decaying particles should be considered stable in the context of coalescence.
As an alternative, the authors of Ref.~\cite{Fornengo:2013osa} implement an explicit condition on the spatial separation between the antinucleons of $\Delta r < 2$~fm in their coalescence model, which in principle is a more correct approach.
However, since most Monte Carlos do not model the spacetime structure resulting from showering and hadronizaton, and we expect very few antideuterons to be produced by decaying final states, we expect the two approaches to be more or less equivalent. 

The coalescence model for antideuteron production describes a $2\rightarrow 1$ process, which does not preserve energy--momentum.
This issue is not much discussed in the literature, but is usually solved by requiring momentum conservation, $\vec p_{\bar d} = \vec p_{\bar p} + \vec p_{\bar n}$, 
and calculating the antideuteron energy through $E_{\bar d} = \sqrt{|\vec p_{\bar d}|^2 + m_{\bar d}^2}$, implicitly assuming that the excess energy is somehow disposed of at a later point.
A more satisfactory description is to consider this as a radiative capture process $\bar p \bar n \rightarrow \bar d \gamma$, which is the dominating antideuteron formation process at the low COM momentum differences required by the coalescence model.
For a full kinematical description, the magnitude and direction of the photon recoil must be taken into account through four-momentum conservation.
However, for antideuteron kinetic energies well above $p_0$, the effect is negligible.
Any spin correlations in the COM system  will affect the angular distributions of the final state particles with respect to the initial state, and should in principle also be taken into account. However, we see no {\it a priori} reason for such a correlation, and the effect will be washed out in the lab-frame by the generally large boost.

\section{An empirical, cross section based model}
\label{sec:Empirical}

\subsection{The model}

In the coalescence model, antideuteron formation is classically deterministic, and the probability that a $\bar p \bar n$-pair will form an antideuteron can be expressed as a step function in the COM momentum difference between the antineutron and antiproton,
\beq
P(\bar p \bar n \rightarrow \bar d\ |\ k) = \theta(p_0 - k),
\eeq
where $k=|\vec{p}_{\bar p} - \vec{p}_{\bar n}|_{\rm COM}$.
From quantum mechanics, one would not expect a relation like this, but rather a formation probability that depends on the wave function overlap of the initial state nucleons, and varies as a function of $k$, just as in an ordinary scattering process.
We expect this probability to be proportional to the cross section for the corresponding capture process $\bar p \bar n \rightarrow \bar d X$,
\beq
P(\bar p \bar n \rightarrow \bar d X\ |\ k) \propto \sigma_{\bar p \bar n \rightarrow \bar d X} (k).
\eeq

As an alternative to the coalescence model, we therefore propose a model in which the combination of a $\bar p \bar n$-pair with COM momentum difference $k$ into an antideuteron is a random event with a probability given by
\beq
P(\bar p \bar n \rightarrow \bar d X\ |\ k) = \frac{\sigma_{\bar p \bar n \rightarrow \bar d X} (k)}{\sigma_0},
\eeq
where $\sigma_{\bar p \bar n \rightarrow \bar d X} (k)$ is the sum of cross sections for $\bar p \bar n$-processes with an antideuteron in the final state, and the constant of proportionality $\sigma_0$ is a free parameter to be fixed through calibration against experimental data, analogous to $p_0$ in the coalescence model.\footnote{While $\sigma_0$ should in principle be calculable, it will in practice depend on properties of the wave-functions of the incoming nucleons.}

For low values of $k$, the relevant process is the radiative capture process $\bar p \bar n \rightarrow \bar d \gamma$.
For COM energies above the pion production threshold, instead processes with hadronic final states $\bar p \bar n \rightarrow \bar d (N \pi)^0$ dominate. 
At these energies, antideuterons are actually more efficiently produced through $\bar p \bar p$ and $\bar n \bar n$ processes with $\bar d (N\pi)$ final states, and these processes must therefore also be taken into account.
The cross sections decrease with increasing number of final states, and experimental data also become significantly more sparse.
In this work, we will as a result only consider the antideuteron production processes listed in Table~\ref{tab:processes}.
\begin{table}[h!]
	\begin{tabular}{c l c l}
	  $1)$& $\bar p \bar n \rightarrow \bar d \gamma$		&\quad\quad\quad\quad$5)$& $\bar p \bar p \rightarrow \bar d \pi^-$		\vspace{2mm} \\
	  $2)$& $\bar p \bar n \rightarrow \bar d \pi^0 $		&\quad\quad\quad\quad$6)$& $\bar p \bar p \rightarrow \bar d \pi^-\pi^0$	\vspace{2mm} \\
	  $3)$& $\bar p \bar n \rightarrow \bar d \pi^+\pi^-$	&\quad\quad\quad\quad$7)$& $\bar n \bar n \rightarrow \bar d \pi^+$		\vspace{2mm} \\
	  $4)$& $\bar p \bar n \rightarrow \bar d \pi^0\pi^0$	&\quad\quad\quad\quad$8)$& $\bar n \bar n \rightarrow \bar d \pi^+\pi^0$	\\
	\end{tabular}
	\caption{Processes considered in this work.}
	\label{tab:processes}
\end{table}

For a given antinucleon pair, the probability that it will form an antideuteron though a process $i$ from Table~\ref{tab:processes} is in our model given by 
\beq \label{eq:crossSecProb}
P(\bar N_1 \bar N_2 \rightarrow \bar d X_i\ |\ k) = \frac{\sigma_{\bar N_1 \bar N_2 \rightarrow \bar d X_i} (k)}{\sigma_0},
\eeq
where $\bar N_1$ and $\bar N_2$ are the species of the two antinucleons, and $X_i$ represents the other final state particles in the given process. 
The free normalization factor $\sigma_0$ is assumed to be the same for all processes. The energy of the produced antideuteron depends on the kinematics of the relevant process, and we will discuss this separately for the different processes in the following sections.

Little or no data is available on the antinucleon processes we consider here, and we will therefore be basing our model on fits to data on the charge conjugate processes under the assumption $\sigma_{\bar N_1 \bar N_2 \rightarrow \bar d X} = \sigma_{N_1 N_2 \rightarrow d \bar X}$.

\subsection{The $\bar p \bar n \rightarrow \bar d \gamma$ process}
We have found only a small amount of data on the $p n \rightarrow d \gamma$ process, and then only at low energies. This alone is not sufficient to make a fit of the cross section as a function of $k$.
However, for the inverse process of photodisintegration, $d \gamma \rightarrow p n$, a large amount of data is available, and can be used through application of the principle of detailed balance --- see {\it e.g.} Ref.~\cite{landau1977quantum} for a detailed description. The principle implies that given time reversal invariance of the interaction, the cross section for a process $\sigma(A a \rightarrow B b)$ is related to the cross section for the inverse process through
\beq
\sigma(A a \rightarrow B b) = \frac{g_B g_b}{g_A g_a} \frac{p_b^2}{p_a^2} \sigma(B b \rightarrow A a),
\eeq
where $p_i$ is the momentum, and $g_i$ is the number of spin states of particle $i$; for massive particles, $g_i = (2s_i + 1)$.
All quantities are given in the COM frame.
Cross sections are invariant under Lorentz boosts along the beam axis, and most experimental cross sections can therefore be used at face value here.
In the derivation of the above expression, applicability of perturbation theory is typically assumed,
but the principle can be shown to be valid also when perturbation theory breaks down, provided that averages over all spin variables have been performed~\cite{belkic2003principles}.

Applying the principle to the process $p n \rightarrow d \gamma$, we have $s_p=s_n=\frac{1}{2}$ and $s_d=1$, giving $g_p=g_n=2$ and $g_d=3$. 
While the photon has spin 1, it is massless and thus only contributes two polarization states, $g_{\gamma}=2$. 
Detailed balance then gives the relation 
\beq
\sigma(p n \rightarrow d \gamma) = \frac{3}{2} \frac{p_{\gamma}^2}{p_{n}^2} \sigma(d \gamma \rightarrow p n),
\eeq
which is frequently used in the literature on radiative capture and deuteron photodisintegration.

Large amounts of experimental data on deuteron photodisintegration can be found in the literature, however,  some of the experiments are in tension with each other. 
In order to be able to make a fit, it is necessary to prune the data down to a consistent dataset. Lacking better information, our approach is therefore to only use data from the most recent experiment in energy ranges where the experiments are in tension.
For consistency, we discard the entire datasets from removed experiments, not only the points that are in tension with other experiments. Our final set of experimental data consists of radiative capture data from Refs.~\cite{Wauters:1990,Stiehler:1985,Suzuki:1995,Nagai:1997zz,Cox:1965} 
and photodisintegration data from Refs.~\cite{Bernabei:1986ai,DeGraeve:1992zz,Birenbaum:1985zz,Arends:1983ae,Myers:1961zz,Crawford:1996ka,Hara:2003gw,Moreh:1989zz}.\footnote{We were unable to reliably extract the errors from Ref.~\cite{Crawford:1996ka}, and instead assumed 5\% errors, which are typical for similar experiments.}

\begin{figure}[h!]
\includegraphics[width=0.5\textwidth]{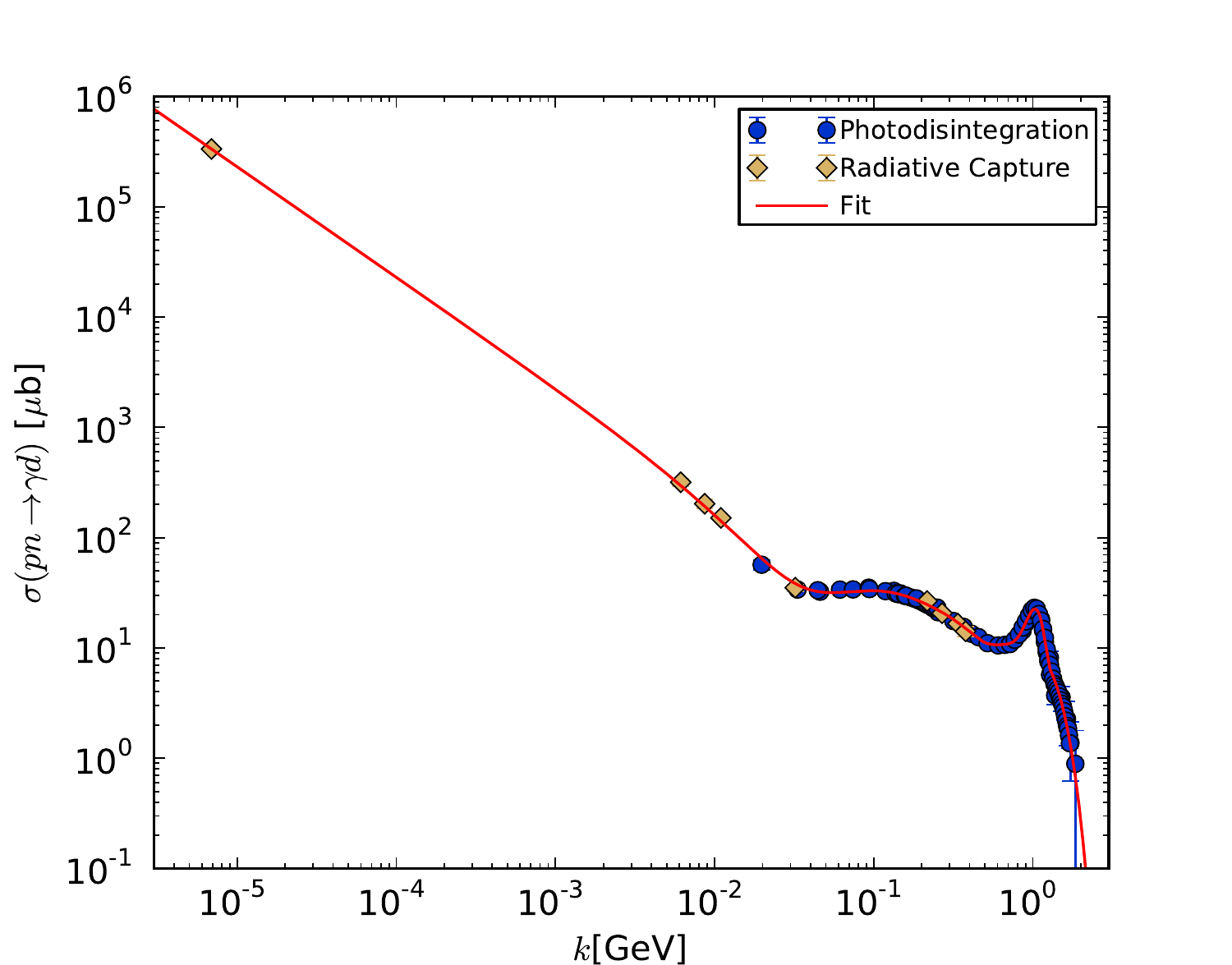}
\caption{Fit to experimental data for the deuteron radiative capture cross section as function of nucleon momentum difference, $k$, in the COM frame. Circles show the photodisintegration data, while diamonds show radiative capture data.}
\label{fig:captureFit_log}
\end{figure}

\begin{figure}[h!]
\includegraphics[width=0.35\textwidth]{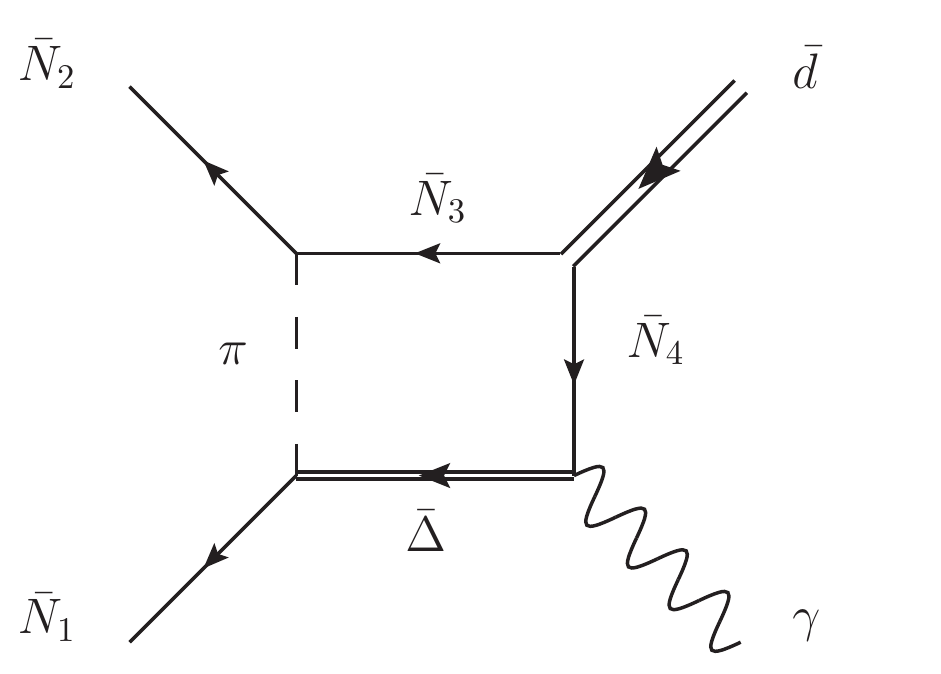}
\caption{Feynman diagram for the delta-resonance in radiative capture. $\bar N_i$ are here antinucleons.}
\label{fig:feyn_delta}
\end{figure}

After applying the principle of detailed balance to translate the photodisintegration data into radiative capture cross sections, we perform a least squares fit to the combined radiative capture data using the function
\beq
  \label{eq:fit}
  \frac{\sigma_{\bar n \bar p \rightarrow \bar d \gamma} (\kappa) }{(1 \mu{\rm b})} =  
  \left\{
  \begin{array}{lr}
	\sum_{n=-1}^{10} a_n \kappa^n	& : \kappa < 1.28 \\
    \exp(-b_1 \kappa -b_2 \kappa^2)	& : \kappa \ge 1.28 ,
  \end{array}
\right.
\eeq
where $\kappa= k / {\rm (1~GeV)}$.
We chose to use an exponential form above $\kappa=1.28$ to ensure that the function does not unphysically diverge or obtain negative values at high energies.
Due to the $\kappa^{-1}$ term, the fit function for the cross section clearly goes to infinity as $k$ approaches 0. 
We therefore take care to restrict the antideuteron production probability to $P(\bar n \bar p \rightarrow \bar d \gamma\ |\ k) \le 1$ when using this fit in Eq.~\eqref{eq:crossSecProb}.
The best fit parameter values for our dataset can be found in Tab.~\ref{tab:fitParams}, and give an excellent fit of $\chi^2 = 51.8$ for 83 degrees of freedom. 
Note that since the fit was made to data spanning 6 orders of magnitude in energy, the parameter values are rather finely tuned, and must therefore be used at the given level of precision.
The experimental data, as well as our fit are plotted as a function of $k$ in Fig.~\ref{fig:captureFit_log}. The peak in the cross section near 1~GeV is due to the delta-resonance --- processes in which one of the nucleons is excited to a delta resonance, via  virtual pion exchange, as seen in Fig.~\ref{fig:feyn_delta}.

\begin{table}
	\begin{tabular}{c r l}
		Parameter 	& 	Value\\ 
		\hline
		$a_{-1}$ 	& 	 2.30346&				\\
		$a_{0}$ 	& 	-9.366346&\e{1}     	\\
		$a_{1}$ 	& 	 2.565390&\e{3}			\\
		$a_{2}$ 	& 	-2.5594101&\e{4}		\\
		$a_{3}$ 	& 	 1.43513109&\e{5}		\\
		$a_{4}$ 	& 	-5.0357289&\e{5}		\\
		$a_{5}$ 	& 	 1.14924802&\e{6}		\\
		$a_{6}$ 	& 	-1.72368391&\e{6}		\\
		$a_{7}$ 	& 	 1.67934876&\e{6}		\\
		$a_{8}$ 	& 	-1.01988855&\e{6}		\\
		$a_{9}$ 	& 	 3.4984035&\e{5}		\\
		$a_{10}$ 	& 	-5.1662760&\e{4}		\\
		$b_{1}$ 	& 	-5.1885&				\\
		$b_{2}$ 	& 	 2.9196&				\\
	\end{tabular}
	\caption{Best fit values to the parameters given in Eq.~\eqref{eq:fit}.}
	\label{tab:fitParams}
\end{table}

In this model, in contrast to the coalescence model, antideuterons can be produced at values of $k$ well into the GeV range, which leaves a potentially large amount of excess energy to be radiated off by the photon. 
This in turn gives a sizeable recoil that must be taken into account by requiring four-momentum conservation. 
In application of the model we let the antideuteron and photon be emitted back-to-back in the COM system, in a random direction drawn from an isotropic distribution.

\subsection{$\bar N_1 \bar N_2 \rightarrow \bar d \pi$ processes}
The $p n \rightarrow d \pi^0$, $n n \rightarrow d \pi^-$ and $p p \rightarrow d \pi^+$ processes are related by isospin invariance through
\beq
\sigma_{p n \rightarrow d \pi^0} = \frac{1}{2} \sigma_{p p \rightarrow d \pi^+},
\eeq
and
\beq
\sigma_{n n \rightarrow d \pi^-} = \sigma_{p p \rightarrow d \pi^+};
\eeq
see {\it e.g.}\ Ref.~\cite{Bystricky:1987}.
These relations are not exact, as the isospin symmetry is broken by the differing nucleon and pion masses. 
Very little data exists for the $p n \rightarrow d \pi^0$ process, and we have not been able to find any data on the $n n \rightarrow d \pi^-$ process.
A substantial amount of data is, however, available on the $p p \rightarrow d \pi^+$ reaction, and we will therefore use these data in combination with the above isospin relations to approximate the $p n \rightarrow d \pi^0$ and $n n \rightarrow d \pi^-$ cross sections.
The authors of Ref.~\cite{Machner:2005ha} have already made a fit to the available data on the $p p \rightarrow d \pi^+$ process, and we will here adopt their fit.
They find the data to be well described by the function
\beq \label{eq:pp_dpiFit}
\sigma(\eta) = \frac{a \eta^b}{(c-\exp(d\eta))^2 +e}
\eeq
with the parameters given in Tab.~\ref{tab:pp_dpiFitParams}, where $\eta = q/m_{\pi^+}$, and $q$ is the momentum of the pion in the COM frame.\footnote{When using the isospin relations, one should for consistency use $m_{\pi^+}$ in calculating $\eta$ also in the $p n \rightarrow d \pi^0$ and $n n \rightarrow d \pi^-$ processes~\cite{Machner:2005ha}.}

\begin{table}[h]
	\begin{tabular}{c c }
		Parameter 	& 	Value \\ 
		\hline
		$a$ [$\mu$b]&	170		\\
		$b$			&	1.34	\\
		$c$			&	1.77	\\
		$d$			&	0.38	\\
		$e$ 		&	0.096	\\
	\end{tabular}
	\caption{Best fit values from Ref.~\cite{Machner:2005ha} to the parameters given in Eq.~\eqref{eq:pp_dpiFit}.}
	\label{tab:pp_dpiFitParams}
\end{table}

The fit was made in the context of comparison to $p n \rightarrow d \pi^0$ data, and was corrected for Coulomb repulsion and phase space differences due to the differing pion and nucleon masses.
These effects should in principle be re-applied to the $pp$-process, and an analogous phase space correction should also be applied when using the fit with the $nn$-process.
However, these effects are only important near threshold for the process, and will effectively shift the threshold slightly in $k$. At high COM energies, the cross section is unchanged, and at the peak the corrections are only at the percent level.
There is no reason to expect the (anti)deuteron spectrum to be sensitive to the precise position of the threshold, so for simplicity we will neglect these corrections here.
We  set the cross sections to zero below the kinematic thresholds for the processes, as this is not ensured by the fit. 

\begin{figure}[h]
\includegraphics[width=0.425\textwidth]{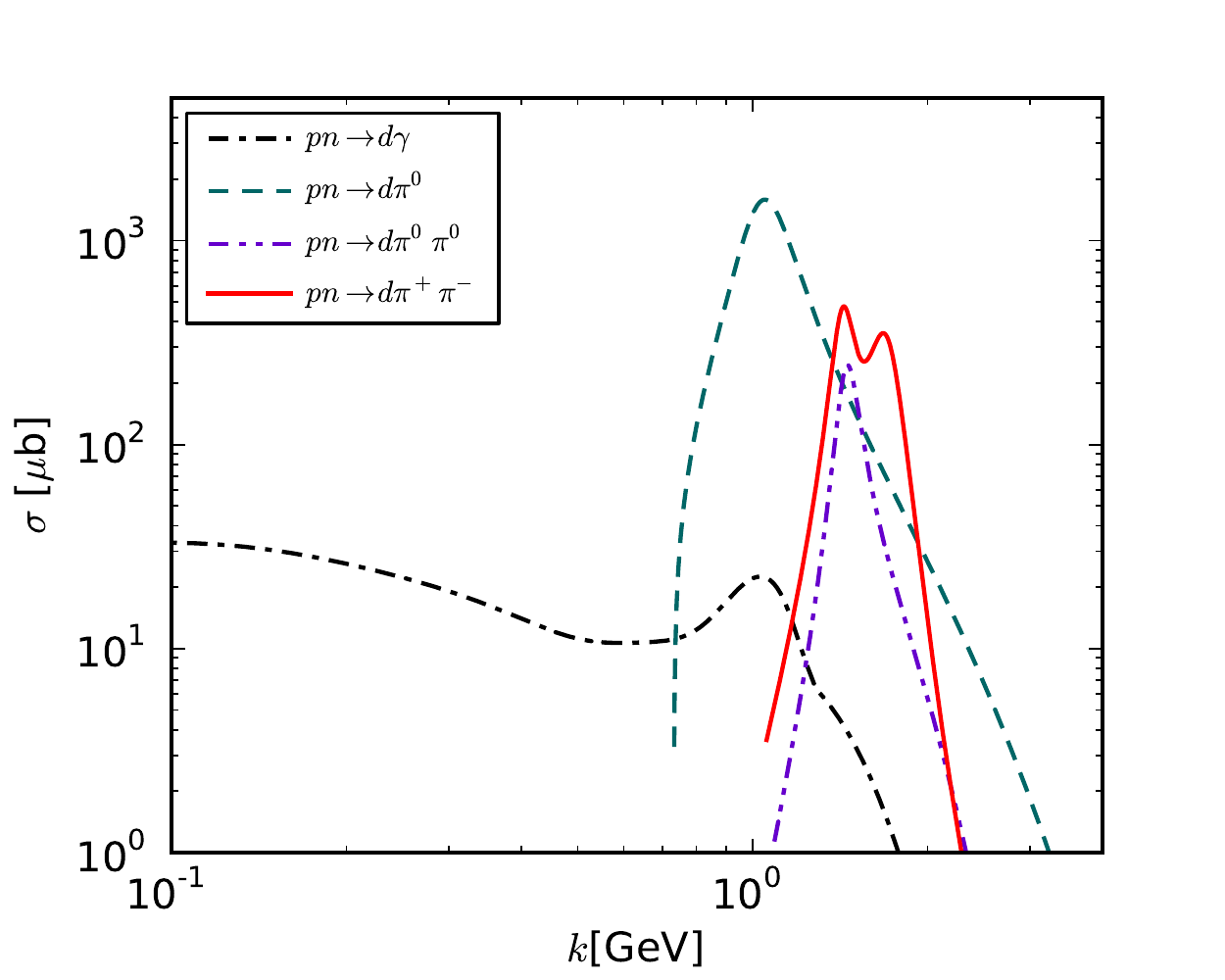}
\caption{Fits to cross sections for $n p \rightarrow d X$ processes.}
\label{fig:pnFits}
\end{figure}

\begin{figure}[h]
\includegraphics[width=0.425\textwidth]{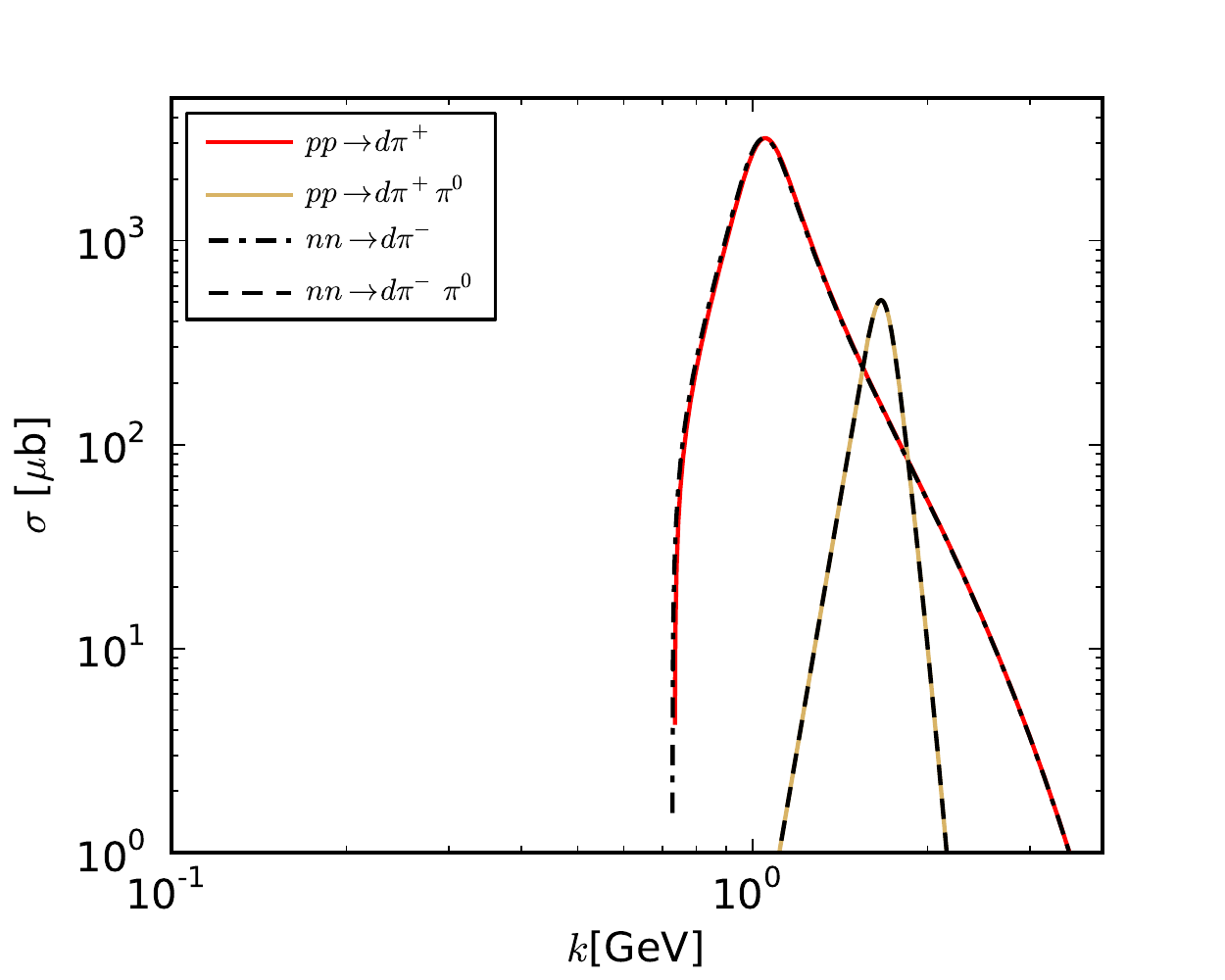}
\caption{Fits to cross sections for $p p \rightarrow d X$ and $n n \rightarrow d X$ processes.}
\label{fig:NNfits}
\end{figure}

We have plotted the cross section fits for the processes as function of the COM momentum difference $k$ in Figs.~\ref{fig:pnFits} and~\ref{fig:NNfits}.
The cross sections for these processes also peak at the delta resonance near $k=1$~GeV, and, just as for the photon in the $\bar p \bar n \rightarrow \bar d \gamma$ case, the pion recoil must be taken into account.
We again emit the antideuteron and pion back-to-back in a random, isotropically drawn direction in the COM frame, and determine the four-momenta from the kinematics.

\subsection{$\bar N_1 \bar N_2 \rightarrow \bar d \pi \bar \pi$ processes} \label{sec:2pi}
For the $\bar N_1 \bar N_2 \rightarrow \bar d \pi \bar \pi$ processes, data are available on all but the $n n \rightarrow d \pi^- \pi^0$ process.
We here use $p p \rightarrow d \pi^+ \pi^0$ data from Refs.~\cite{Kren:2009hr,Shimizu:1982dx,Bystricky:1987,Adlarson:2012fe}, $p n \rightarrow d \pi^+ \pi^-$ data from Refs.~\cite{Bystricky:1987,Adlarson:2012fe,BarNir:1973wb}, and $p n \rightarrow d \pi^0 \pi^0$ data from Refs.~\cite{Adlarson:2012fe,Adlarson:2011bh}.\footnote{Many of the datasets are only available as plots, and for cases where errorbars cannot be resolved, we use the point size of the plot as an estimate for the error.}
There is unfortunately very little data available for $\sqrt{s} > 2.5$~GeV for all the processes. This makes fits to the $p p \rightarrow d \pi^+ \pi^0$ and $p n \rightarrow d \pi^+ \pi^-$ processes particularly problematic.
The exact locations and heights of the resonance peaks near $\sqrt{s} = 2.5$~GeV in these two processes are unclear, and for the $p p \rightarrow d \pi^+ \pi^0$ process, the lack of data at high energies makes the na\"ive fit quite unstable.
To improve on this, we again make use of isospin invariance.

Isospin invariance predicts the relations~\cite{Bystricky:1987}
\beq \label{eq:isospin_2pi}
\sigma_{p n \rightarrow d \pi^+ \pi^-} = 2 \sigma_{p n \rightarrow d \pi^0 \pi^0} + \frac{1}{2} \sigma_{p p \rightarrow d \pi^+ \pi^0},
\eeq
and
\beq \label{eq:isospin_2pi_nn}
\sigma_{n n \rightarrow d \pi^- \pi^0} = \sigma_{p p \rightarrow d \pi^+ \pi^0},
\eeq
between the cross sections.
Measurements of the processes in Eq.~\eqref{eq:isospin_2pi} within the same experiment~\cite{Adlarson:2012fe} have shown these cross sections to be quite sensitive to isospin breaking effects, leading to a $\sim 25\%$ deviation in this relation. 
If the isospin symmetry was exact, one could have used Eq.~\eqref{eq:isospin_2pi} to make simultaneous fits to all three processes, but due to the isospin breaking we have not been able to obtain good fits in this manner.
We therefore instead perform individual fits to each process, where we include the data from the other processes through Eq.~\eqref{eq:isospin_2pi} for stability, but weighted down by a factor 1/100 in the $\chi^2$.
We have chosen the value of the weight to be large enough to guide the fits, giving reasonable high energy behaviour in the $p p \rightarrow d \pi^+ \pi^0$ and $p n \rightarrow d \pi^+ \pi^-$ channels, but low enough to give good individual fits for the different processes.
To further guide the fits, we also insert dummy data points at the kinematic cutoffs for the processes, with zero cross section, and errors of 1~$\mu$b.

\begin{figure*}[ht]
\includegraphics[width=0.3\textwidth]{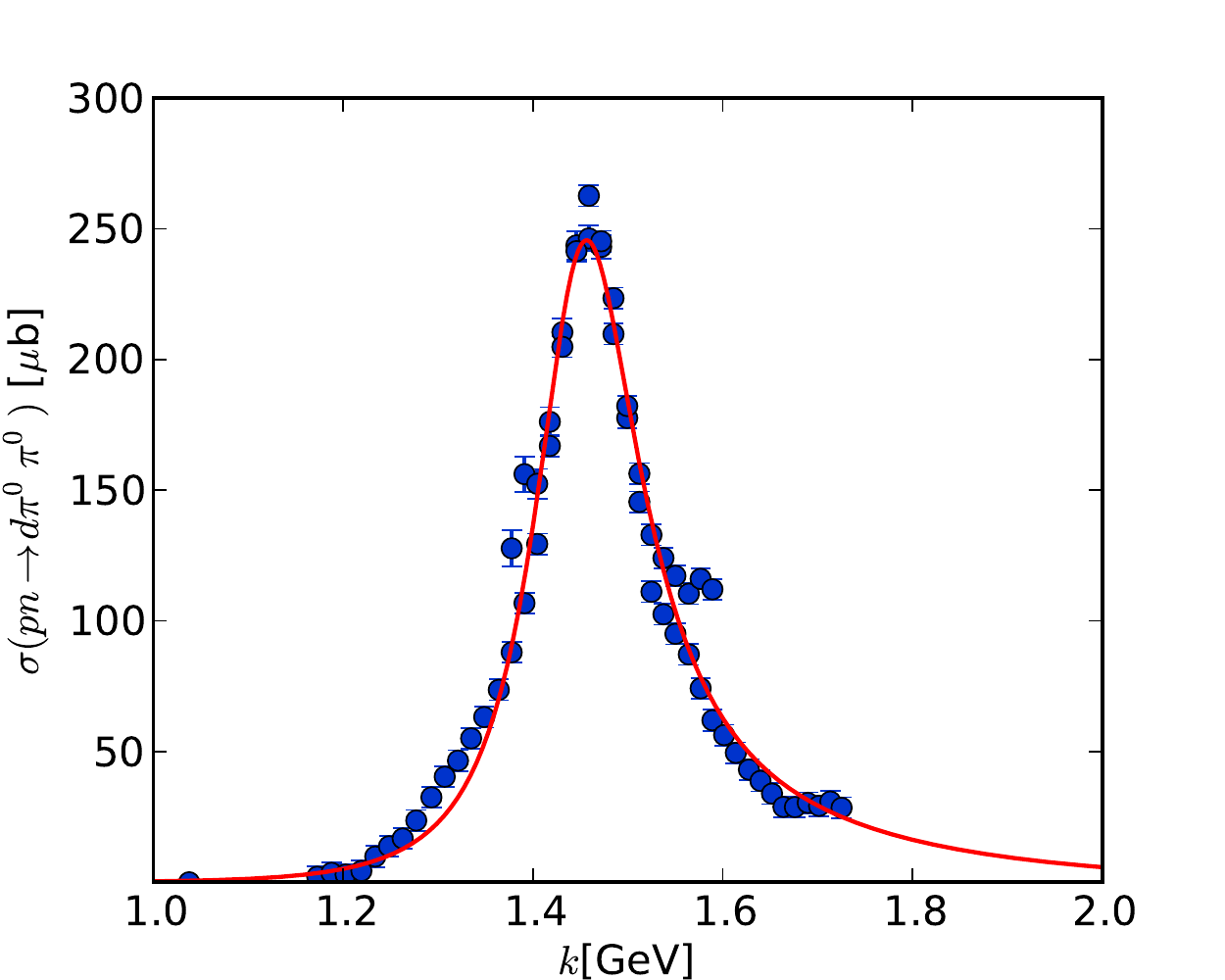}
\includegraphics[width=0.3\textwidth]{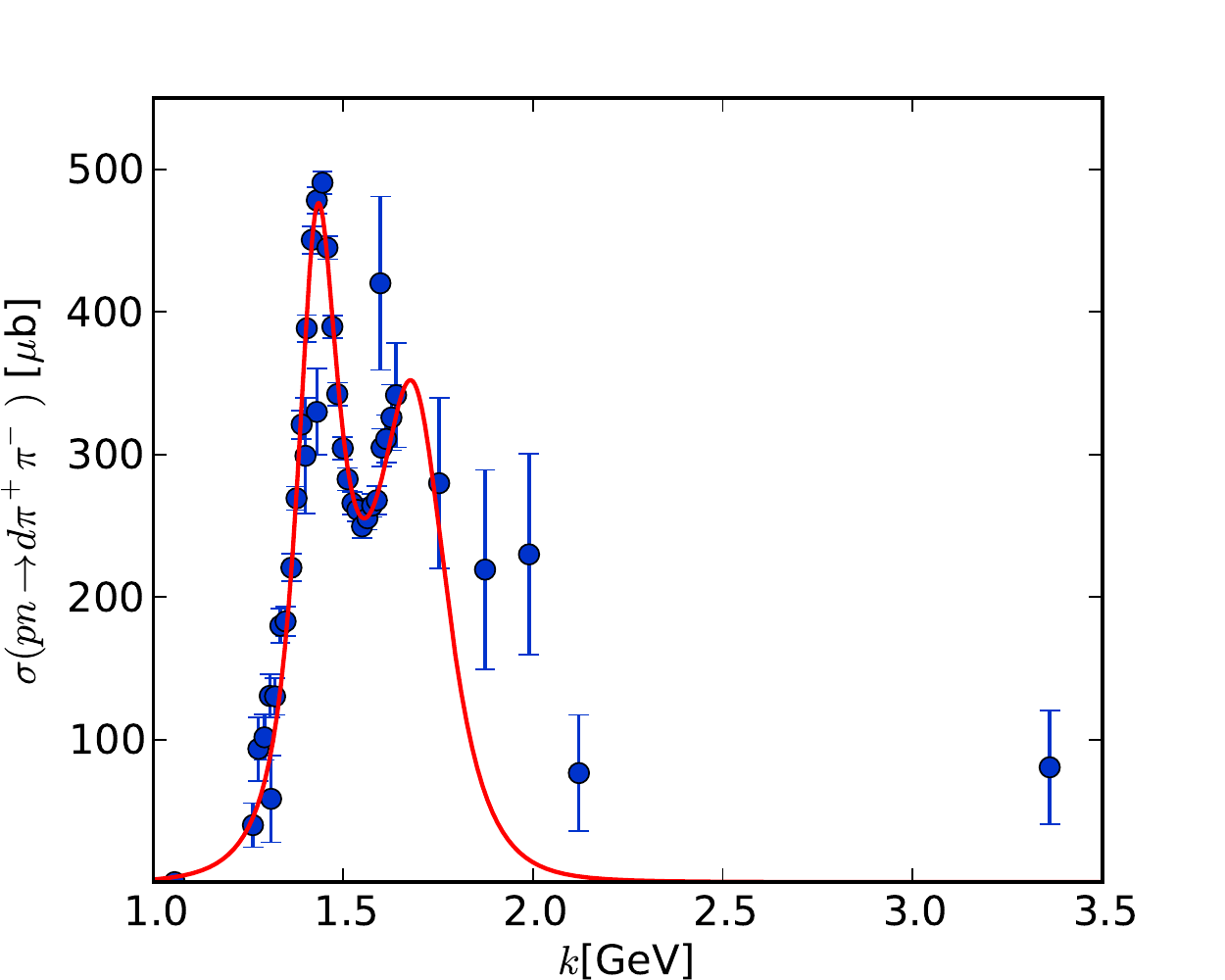}
\includegraphics[width=0.3\textwidth]{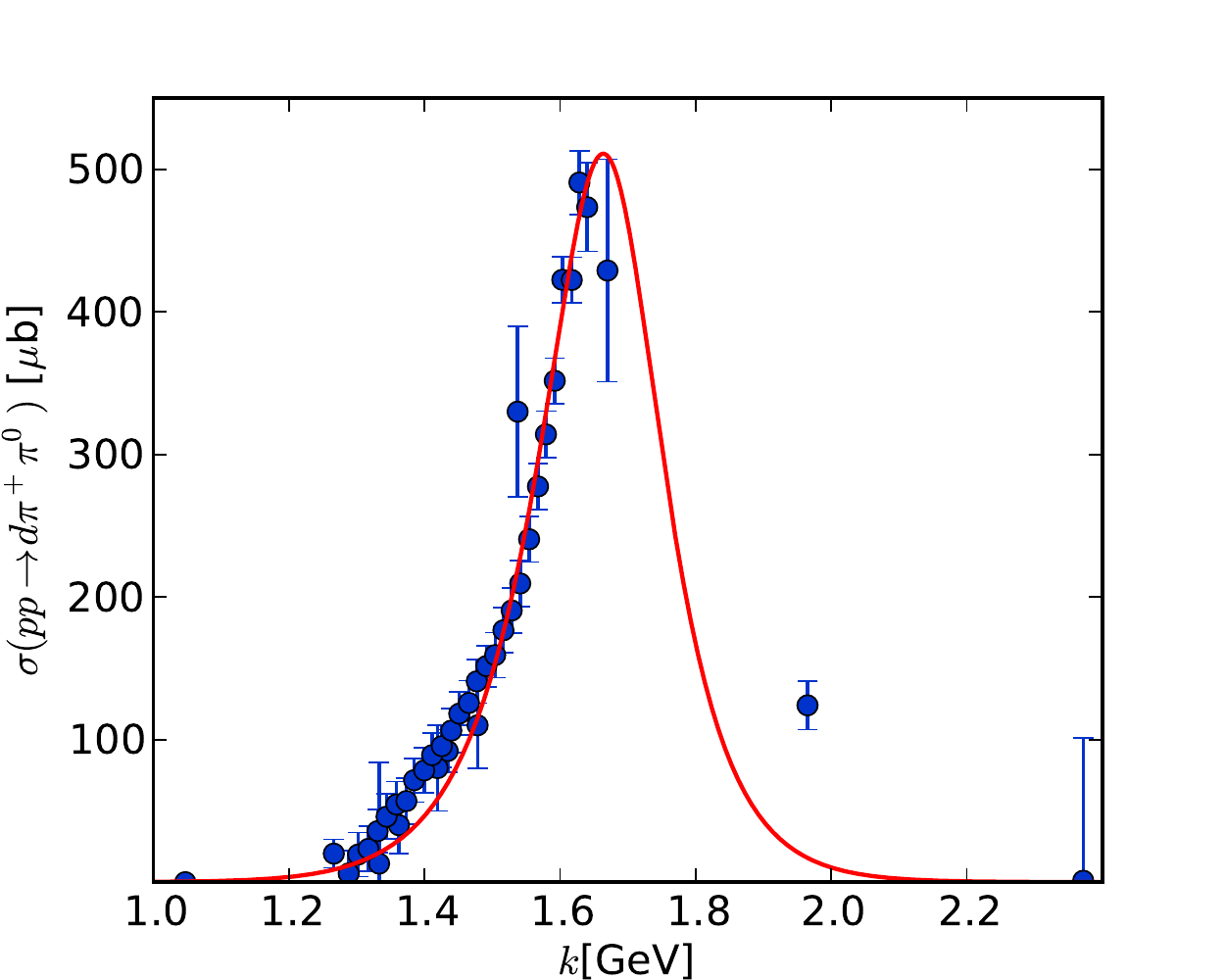}
\caption{Fits to cross section data. Left: $p n \rightarrow d \pi^0 \pi^0$, middle: $p n \rightarrow d \pi^+ \pi^-$, right: $p p \rightarrow d \pi^+ \pi^0$.}
\label{fig:2piFits}
\end{figure*}

We use the following functional forms for the fits, inspired by~\cite{Machner:2005ha},
\beq \label{eq:FitFunc}
\sigma(\kappa) = \frac{a \kappa^b}{(c-\exp(d \kappa))^2 +e},
\eeq
for the $p p \rightarrow d \pi^+ \pi^0$ and $p n \rightarrow d \pi^0 \pi^0$ processes, and
\beq \label{eq:FitFunc2}
\sigma(\kappa) = \frac{a_1 \kappa^{b_1}}{(c_1-\exp(d_1 \kappa))^2 +e_1} + \frac{a_2 \kappa^{b_2}}{(c_2-\exp(d_2 \kappa))^2 +e_2},
\eeq
for $p n \rightarrow d \pi^+ \pi^-$, where again $\kappa = k /  ({\rm 1~GeV})$.
The best fit parameters for the different processes are listed in Tables~\ref{tab:pn_dpi0pi0FitParams}, \ref{tab:pn_dpippimFitParams}, and~\ref{tab:pp_dpippi0FitParams}.
The data points used and our fits to these points are plotted as functions of $k$ in Fig.~\ref{fig:2piFits}.
No data are available on the  $n n \rightarrow d \pi^- \pi^0$ process, and we are forced to make use of Eq.~\eqref{eq:isospin_2pi_nn} to approximate the cross section for this process.
As in the $\bar N_1 \bar N_2 \rightarrow \bar d \pi$ case, we set the cross sections to zero below the kinematic thresholds. 

\begin{table}[h]
	\begin{tabular}{c c }
		Parameter 	& 	Value \\ 
		\hline
		$a$ [$\mu$b]&	2.855\e{6}	\\
		$b$			&	1.311\e{1}	\\
		$c$			&	2.961\e{3}	\\
		$d$			&	5.572\e{0}	\\
		$e$ 		&	1.461\e{6}	\\
	\end{tabular}
	\caption{Best fit parameters for the $p n \rightarrow d \pi^0 \pi^0$ process.}
	\label{tab:pn_dpi0pi0FitParams}
\end{table}

\begin{table}[h]
	\begin{tabular}{c c}
		Parameter 	& 	Value\\ 
		\hline
		$a_1$ [$\mu$b]	&	6.465\e{6}	\\	 
		$b_1$			&	1.051\e{1}	\\ 
		$c_1$			&	1.979\e{3}	\\ 
		$d_1$			&	5.363\e{0}	\\ 
		$e_1$			&	6.045\e{5}	\\
		$a_2$  [$\mu$b]	&	2.549\e{15} \\
		$b_2$			&	1.657\e{1} 	\\
		$c_2$			&	2.330\e{7}	\\ 
		$d_2$			&	1.119\e{1} 	\\
		$e_2$			&	2.868\e{16}	\\
	\end{tabular}
	\caption{Best fit parameters for the $p n \rightarrow d \pi^+ \pi^-$ process.}
	\label{tab:pn_dpippimFitParams}
\end{table}

\begin{table}[h]
	\begin{tabular}{c c}
		Parameter 	& 	Value \\ 
		\hline
		$a$ [$\mu$b]&	5.099\e{15} \\
		$b$			&	1.656\e{1} 	\\
		$c$			&	2.333\e{7} 	\\
		$d$			&	1.133\e{1} 	\\
		$e$			&	2.868\e{16}	\\
	\end{tabular}
	\caption{Best fit parameters for the $p p \rightarrow d \pi^+ \pi^0$ process.}
	\label{tab:pp_dpippi0FitParams}
\end{table}

As these processes have three-body final states, the kinematics become considerably more involved than in the previous cases.
For a detailed review, we refer to the section on three-body decays in Ref.~\cite{Agashe:2014kda}.
In processes with three-body final states, there can be angular correlations between the final states that depend on the matrix element for the process, and this is the case for the processes considered here.
Dalitz plots from measurements of the $pn$ and $pp$ processes can be found for a few different COM energies in Ref.~\cite{Adlarson:2012fe}, but these data are not sufficient to parameterize the deuteron COM momentum distribution as function of energy.
We therefore make the approximation of no angular correlations between the outgoing deuteron and pions, and draw the deuteron momentum based on phase space alone.
We determine the deuteron momentum by first drawing random invariant masses $m_{\pi \pi}^2$ and $m_{d \pi}^2$ uniformly within the kinematically allowed region. The momentum of the deuteron in the COM frame is then given by
\beq
  p_d = \sqrt{ \left(\frac{s+m_d^2-m_{d\pi}^2}{2\sqrt{s}} \right)^2 - m_d^2},
\eeq
and we draw its direction from an isotropic distribution in the COM frame.

\subsection{Process contributions}
\begin{figure*}[h!t]
\includegraphics[width=0.325\textwidth]{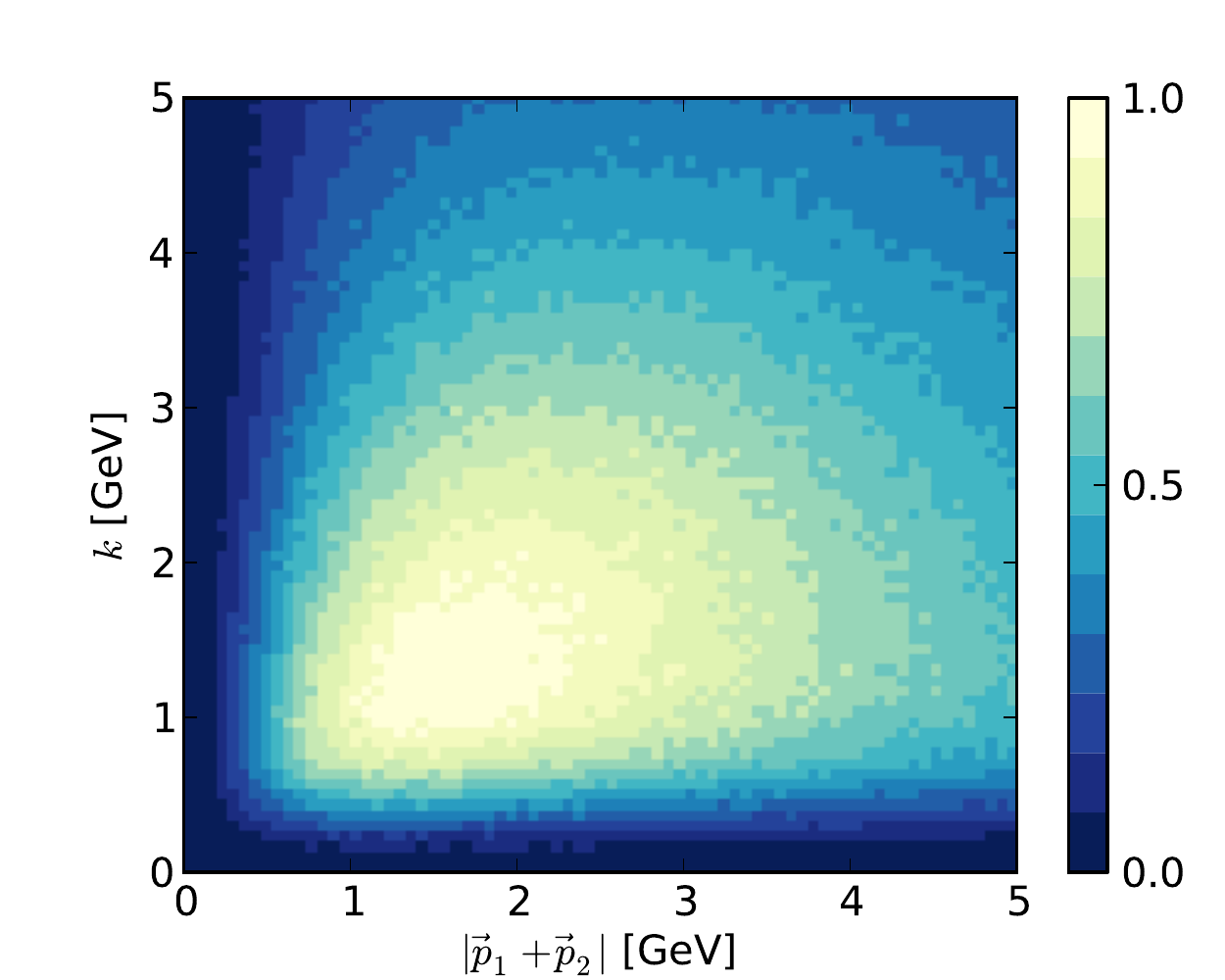}
\includegraphics[width=0.325\textwidth]{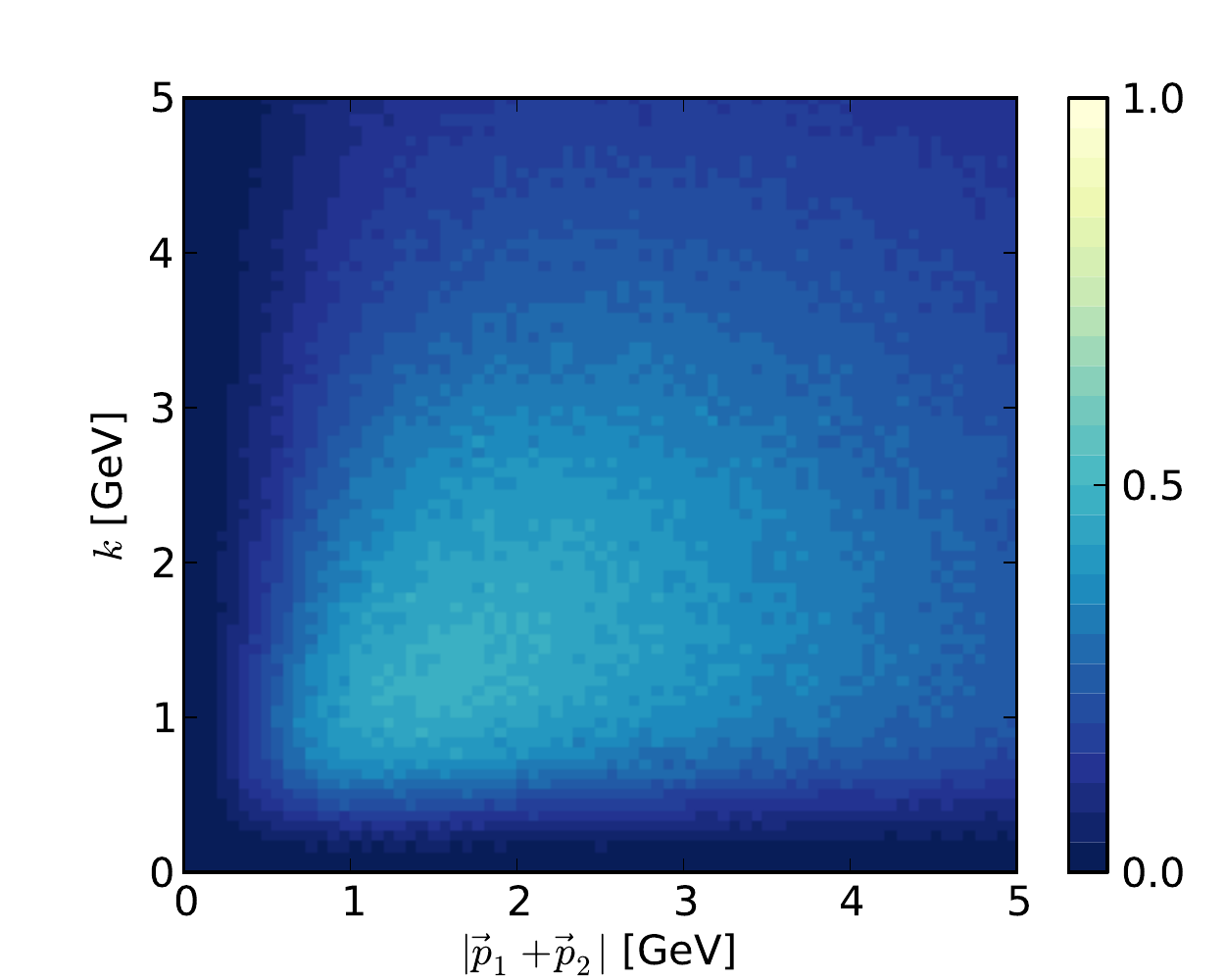}
\includegraphics[width=0.325\textwidth]{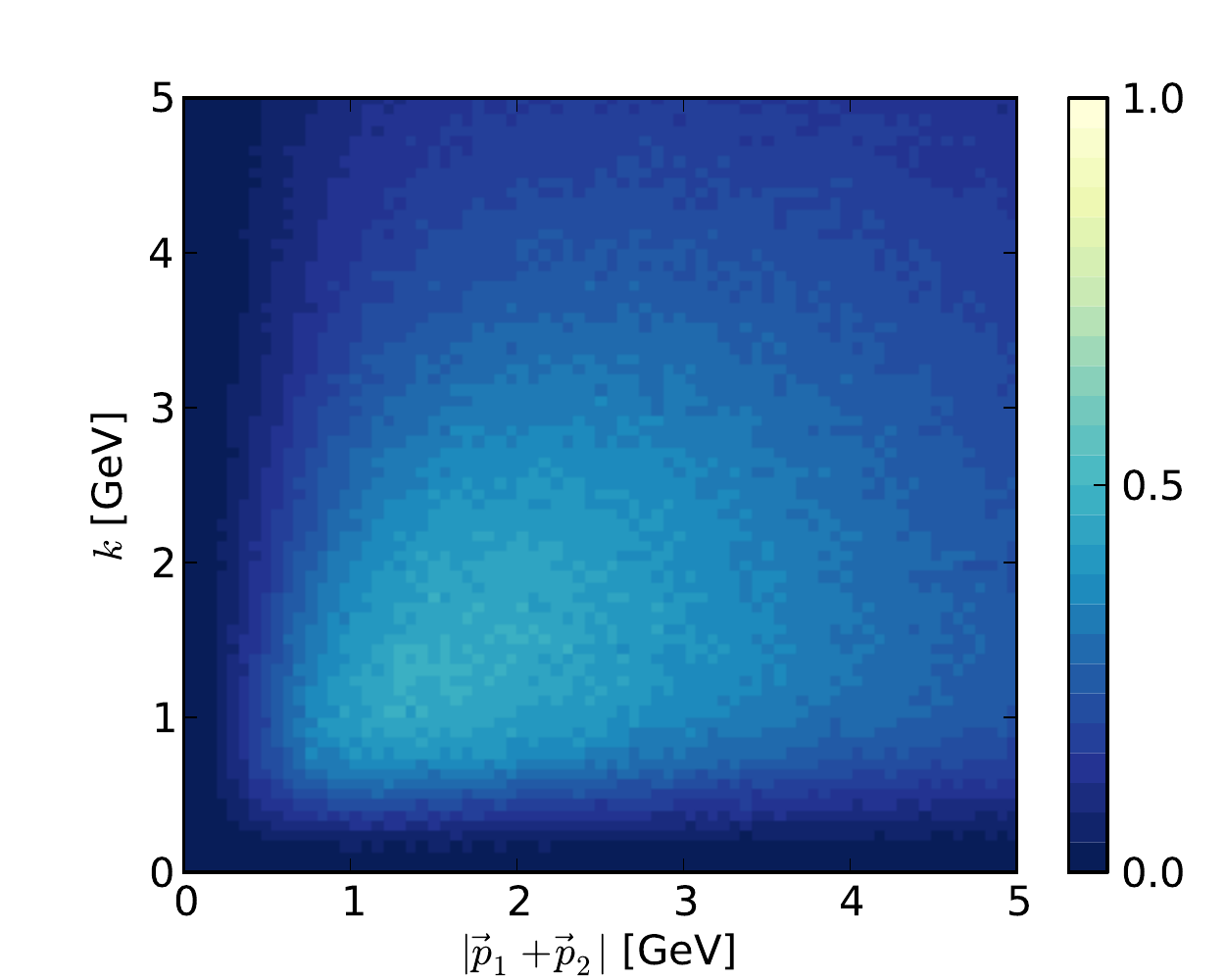}
\caption{Number of available antinucleon pairs in $e^+e^-$-collisions at the $Z$-resonance in {\tt Herwig++}, as a function of antinucleon momentum difference $k$ in the COM frame and total momentum of the pairs in the lab frame. Left: $\bar p \bar n$-pairs, middle: $\bar p \bar p$-pairs, right: $\bar n \bar n$-pairs. The plots have a shared normalization, and are normalized to a maximum value of 1.}
\label{fig:pairDist}
\end{figure*}

In the coalescence model, all antideuterons are by construction produced by $\bar p \bar n$-pairs with low COM momentum differences.
Our model, on the other hand, has the majority of antideuterons produced close to the delta resonance near $k=1$~GeV.
While radiative capture $\bar p \bar n \rightarrow \bar d \gamma$ has a very high cross section at low values of $k$, the number of available $\bar p \bar n$ pairs drops very quickly for decreasing values of $k$ in the processes we have studied.
This can be seen in Fig.~\ref{fig:pairDist}, where we show the number of possible antinucleon--antinucleon combinations in LEP events at the $Z$-peak, generated using {\tt Herwig++ 2.6.0}, as function of $k$ and the combined momenta of the antinucleon pairs in the lab frame.
The distribution peaks for values of $k$ in the low GeV range and drops quickly for decreasing values of $k$. 
The result is that radiative capture at low values of $k$ gives a very small contribution to the total antideuteron spectrum in the cross section based model. Instead, the peak in the number of antinucleon pairs is close to the delta resonance for all values of the combined lab frame momentum.
Antideuteron production is thus dominated by the $\bar N_1 \bar N_2 \rightarrow \bar d (N\pi)$ processes for more or less any antideuteron lab-frame momentum\footnote{The total momentum of an antinucleon pair is an approximation for the momentum of the resulting antideuteron.}. This holds true in all the experiments we consider in this work.

Another notable feature in Fig.~\ref{fig:pairDist}, is that the total number of available $\bar p \bar n$-pairs is roughly a factor two larger than the corresponding numbers of $\bar p \bar p$ and $\bar n \bar n$ pairs, as can be expected from pure combinatorics.
In LEP events, antiprotons and antineutrons are produced with approximately equal probabilities. 
Picking two random antinucleons that each have equal probability of being an antiproton or antineutron is twice as likely to give a $\bar p \bar n$-pair than it is to give either a $\bar p \bar p$-pair or a $\bar n \bar n$-pair.
Since the cross sections for antideuteron production in $\bar p \bar p$ and $\bar n \bar n$ processes are a factor two larger than the $\bar p \bar n$ cross section in the delta resonance region, this implies that $\bar p \bar n$, $\bar p \bar p$ and $\bar n \bar n$ processes give similar contributions to the antideuteron spectrum in our model.

\subsection{Monte Carlo implementation} \label{sec:procedure}
The following is a step-by-step description of how our model should be applied to Monte Carlo events:
\begin{itemize}
\item For each event, iterate over all possible unique antinucleon--antinucleon pairs,  avoiding double counting $\bar p \bar p$ and $\bar n \bar n$-pairs.
\item For each antinucleon-pair in the event, calculate the momentum difference $k$ between the antinucleons in their COM frame. Calculate the probabilities $P(\bar N_1 \bar N_2 \rightarrow \bar d X_i\ |\ k)$ that the antinucleon pair will form an antideuteron for each relevant processes $i$ listed in Table~\ref{tab:processes} using Eq.~\eqref{eq:crossSecProb}. 
	   $\sigma_{\bar N_1 \bar N_2 \rightarrow \bar d X_i} (k)$ is given by the fits in the previous sections,\footnote{Note that the cross sections for the processes with a single pion in the final state are parameterized on $\eta= q / m_{\pi^+}$ (where $q$ is the pion momentum in the COM frame), rather than $k$.} and $\sigma_0$ has to be determined by fits against experimental data.\footnote{We will return to the fitted $\sigma_0$ values below.}
 \item Draw a random number $r_i$ uniformly on the unit interval for each possible formation process. If $r_i<P(\bar N_1 \bar N_2 \rightarrow \bar d X_i\ |\ k)$ for one of the processes, the antinucleon pair forms an antideuteron 			  through that process. If $r_i<P(\bar N_1 \bar N_2 \rightarrow \bar d X_i\ |\ k)$ for more than one process, pick one of the processes randomly using probabilities equal to the relative cross sections.\footnote{As the 		  probability of having multiple successful processes is very low, the way in which a process is chosen in these cases has no significant effect on the final spectrum.}
	   If an antideuteron is formed, exclude the involved antinucleons from being used in the formation of other antideuterons.\footnote{The authors of~\cite{Grefe:2015jva} estimate that even at large values of $p_0=250$ MeV in the coalescence model multiple successful antideuteron candidates are found in less than 0.1\% of the events, and are thus a negligible problem. This also holds true in our model due to the low probabilities for each pair. It is interesting to speculate if these very rare events could be used to estimate the production of even heavier antinuclei such as $^3\overline{\rm He}$.}
 \item Emit the antideuteron in a random, isotropically drawn direction in the COM frame. For two-body final states, its energy and momentum is determined by four-momentum conservation. 
	   For three-body final states, draw the antideuteron momentum randomly based on the available phase space, as discussed in Sec.~\ref{sec:2pi}.
 \item Boost the antideuteron to the lab frame.
\end{itemize}

\subsection{Extracting more information}
In the cross section based approach, the determination of whether or not an antinucleon--antinucleon pair will form an antideuteron is probabilistic.
In a single event, there can be many possible $\bar N \bar N$-combinations, each with a non-zero probability to produce an antideuteron.
These probabilities will typically be very low, and in most cases, none of the combinations will produce antideuterons --- the event has essentially gone to waste in the Monte Carlo statistics.
Even in events where antideuterons are produced, other $\bar N \bar N$-combinations could also have been possible, and this information would remain unused. This situation is similar to the one in the coalescence model, where extremely large statistics are needed in order to get a precise antideuteron spectrum.

Moreover, for a given antinucleon pair, the energy of the resulting antideuteron is not fixed: since the COM frame of the antinucleon pair typically will be boosted, the antideuteron energy will be determined by the randomly chosen direction in which it is emitted.
There is, in other words, much more information in each event than will be extracted by a single application of the antideuteron production model.

In order to extract more information from these events, one can use weighted antideuteron events.
For each Monte Carlo event we:
\begin{itemize}
 \item Set up a temporary histogram with the same binning as the one used for the total antideuteron spectrum (main histogram).
 \item If the event contains more than one antinucleon, evaluate the event $N_{\rm samp}$ times, following the procedure in Sec.~\ref{sec:procedure}, and adding any antideuterons to the temporary histogram. 
\end{itemize}
For a given bin, $b$, in the main histogram, the number of antideuterons can then be calculated as
\beq
 N^{\bar d}_b = \sum_{i=1}^{N_{\rm MC}} w_{b,i},
\eeq
where $N_{\rm MC}$ is the total number of Monte Carlo events,
\beq
 w_{b,i} = \frac{ N^{\bar d}_{b,i} }{ N_{\rm samp} },
\eeq
is the contribution to this bin from Monte Carlo event $i$, and $N^{\bar d}_{b,i}$ is here the number of antideuterons in bin $b$ in the temporary histogram of event $i$ after $N_{\rm samp}$ evaluations.
The error on the number of antideuterons in bin $b$ in the main histogram is then given by
\beq \label{eq:weightedSigma}
 \sigma_b = \sqrt{ \sum_{i=1}^{N_{\rm MC}} w_{b,i}^2}.
\eeq
This error is found under the assumption that $N_{\rm samp}$ is large enough to give a representative sample of the antideuteron spectrum in each event, 
but in practice we find it to give a good error estimate even with a relatively low number of $N_{\rm samp} = 10$.
Using this method, we have found that it is possible to achieve the same level of precision with an order of magnitude fewer Monte Carlo events. This would not be possible in the coalescence model, as the antideuteron formation in that model is deterministic, and no more information could be gained by evaluating the same event multiple times.

\subsection{An historical aside}
Modeling deuteron production based on experimentally measured nucleon-nucleon cross sections was discussed in the original coalescence paper by Schwarzschild and Zupan\v{c}i\v{c}~\cite{Schwarzschild:1963zz} from 1963, but they found this approach to yield too few antideuterons.
They thus argued for the presence of a mechanism in which interactions with the surrounding nuclear matter affects the production of deuterons, and introduced the coalescence model as a phenomenological, simplified version of a model proposed by Butler and Pearson~\cite{Butler:1961pr}.

This approach was criticized by Kamal {\it et al.}\ in an article from 1966~\cite{Kamal:1966}. Here, they point out that the $N N \rightarrow d \pi$ processes have a resonant behaviour, and that Schwarzschild and Zupan\v{c}i\v{c} had significantly underestimated the cross section for these processes.
They also note the necessity of including contributions from $p p \rightarrow d \pi^+$ and $n n \rightarrow d \pi^-$ processes.
Taking the resonant behaviour and the extra processes into account, they obtained results in agreement with the deuteron production in their own experiment, and thus rejected the arguments for the introduction of the coalescence model.
This controversy has apparently been more or less forgotten, and the coalescence model has remained state-of-the-art up to today.

Experimental cross sections have also been used to estimate deuteron production in later works, such as Ref.~\cite{BarNir:1973wb}. Here, the authors model the deuteron formation in the $n p \rightarrow d \pi^+ \pi^-$ process as a $n p \rightarrow N N \pi$ process followed by a $N N \rightarrow d \pi$ process.
They use a conventional scattering model to describe the first step of the process, and then use experimentally measured cross sections for the $N N \rightarrow d \pi$ processes to model the deuteron formation. Using this model, they obtain results in reasonable agreement with the experimental measurements.
\footnote{After completion of this work, we have also been made aware of a paper by Gugelot and Paul~\cite{Gugelot:1993hx} from 1993 that suggested a cross section based deuteron formation model similar to ours for use in Monte Carlos. 
At the time, there was unfortunately not sufficient experimental data available to properly test the model, and the work has largely gone unnoticed.  We would like to thank Sebastian Wild for notifying us of this work.}

\section{Calibration against experimental data}
\label{sec:Calibration}
The coalescence model and the cross section based model each have a free parameter that needs to be tuned against experimental data. 
We here present the best fit parameter values for various experiments and two different Monte Carlo event generators, giving necessary details on the experiments and event generation for reproducibility. 
We use the {\tt Herwig++~2.6.0}~\cite{Bahr:2008pv,Arnold:2012fq}  and {\tt Pythia~8.186}~\cite{Sjostrand:2006za,Sjostrand:2007gs} event generators with default settings, unless stated otherwise.
In order to be able to compare the fits for the two models, we only use the experimental uncertainty in calculating $\chi^2$ for the fits. This is to avoid bias in the $\chi^2$ due to differing statistics from the event generation. The statistical uncertainty is in all cases small relative to the experimental uncertainty, so the effect of this should be small.
The resulting values for $p_0$ and $\sigma_0$ are listed in Tables~\ref{tab:bestFitHpp} and~\ref{tab:bestFitPy8}.
As the antideuteron spectrum in the cross section based model scales as $\propto 1/\sigma_0$, we present the results for this model in terms of $1/\sigma_0$, rather than $\sigma_0$ itself.

As discussed in Sec.~\ref{sec:Coalescence}, antinucleons from weak decays should be excluded in the context of antideuteron production, and we thus set all particles with mean lifetimes above 100~fm/c to be stable.\footnote{For the CLEO experiment, we set the limit at 1\AA/c to allow antideuterons in $\Upsilon(1S)$ decays which is the focus of that experiment.}

\begin{table*}
\begin{tabular}{l c c c c c c c c c}
	Experiment 						& Data points	&\quad& Best fit $p_0$ [MeV]	&\quad& $\chi^2_{p_0}$ &\quad& Best fit $1/\sigma_0$ [$\rm barn^{-1}$]	&\quad& $\chi^2_{\sigma_0}$	\\ 
    \hline
	d,      ALICE, 0.9 TeV 			&	3			&&		228			&&	1.60	&&	5.85	&&	0.71	\\
	$\rm \bar d$, ALICE, 0.9 TeV	&	3			&&		229			&&	7.53	&&	6.15	&&	5.88	\\
	d,      ALICE, 2.76 TeV 		&	7			&&		199			&&	39.8	&&	4.03	&&	10.9	\\
	$\rm \bar d$, ALICE, 2.76 TeV 	&	7			&&		200			&&	74.4	&&	4.00	&&	25.3	\\
	d,      ALICE, 7 TeV 			&	20			&&		181			&&	1001	&&	3.35	&&	231		\\
	$\rm \bar d$, ALICE, 7 TeV 		&	20			&&		185			&&	488		&&	3.40	&&	97.4	\\
	$\rm \bar d$, BABAR 			&	9			&&		94			&&	10.6	&&	0.63	&&	9.01	\\
	$\rm \bar d$, CERN ISR			&	4+4			&&		274			&&	5.15	&&	9.00	&&	5.90	\\
	$\rm \bar d$, CLEO				&	5			&&		130			&&	7.04	&&	0.90	&&	2.11	\\
	$\rm \bar d$, LEP				&	1+1			&&		152			&&	3.61	&&	1.93	&&	3.53	\\
\end{tabular}
\caption{Best fit parameters for the coalescence model and the cross section model in {\tt Herwig++}.}
\label{tab:bestFitHpp}
\end{table*}

\begin{table*}
\begin{tabular}{l c c c c c c c c c}
	Experiment 						& Data points	&\quad& Best fit $p_0$ [MeV]	&\quad& $\chi^2_{p_0}$ &\quad& Best fit $1/\sigma_0$ [$\rm barn^{-1}$]	&\quad& $\chi^2_{\sigma_0}$	\\ 
    \hline
	d,      ALICE, 0.9 TeV 			&	3			&&		201			&&	3.16	&&	3.58	&&	0.77	\\
	$\rm \bar d$, ALICE, 0.9 TeV	&	3			&&		201			&&	8.84	&&	3.63	&&	5.35	\\
	d,      ALICE, 2.76 TeV 		&	7			&&		194			&&	23.7	&&	2.93	&&	9.22	\\
	$\rm \bar d$, ALICE, 2.76 TeV 	&	7			&&		196			&&	46.4	&&	2.88	&&	14.1	\\
	d,      ALICE, 7 TeV 			&	20			&&		194			&&	344		&&	2.63	&&	55.1	\\
	$\rm \bar d$, ALICE, 7 TeV 		&	20			&&		195			&&	113		&&	2.58	&&	12.7	\\
	$\rm \bar d$, BABAR 			&	9			&&		145			&&	16.8	&&	1.13	&&	10.1	\\
	$\rm \bar d$, CERN ISR			&	4+4			&&		151			&&	2.72	&&	2.08	&&	3.26	\\
	$\rm \bar d$, CLEO				&	5			&&		133			&&	1.16	&&	1.25	&&	1.32	\\
	$\rm \bar d$, LEP				&	1+1			&&		183			&&	3.27	&&	1.80	&&	3.55	\\
\end{tabular}
\caption{Best fit parameters for the coalescence model and the cross section model in {\tt Pythia~8}.}
\label{tab:bestFitPy8}
\end{table*}

\subsection{ALICE}
Deuteron and antideuteron spectra in $p\bar p$ minimum bias events at 0.9, 2.76 and 7~TeV have been measured by the ALICE experiment at the LHC~\cite{Serradilla:2013yda}. The ALICE data are particularly interesting, as we here have measurements of both deuteron and antideuteron yields at different energies within the same experiment.
Since Monte Carlo antideuteron production in the coalescence model has shown a strong dependence on the process, 
there has been some speculation as to whether or not the coalescence model can reproduce both the deuteron and antideuteron spectrum in a single experiment with the same value of $p_0$. While, as can be glanced from Tables~\ref{tab:bestFitHpp} and~\ref{tab:bestFitPy8}, the coalescence model yields poor fits to the high-energy data, the best fit values of $p_0$ (and $\sigma_0$) are consistent between deuterons and antideuterons at the different energies in both Monte Carlos. The implication is that any future deuteron data will be very valuable for testing our model, or any other model of antideuteron formation.

In addition to a minimum bias selection, the ALICE analysis uses a trigger (V0AND), which suppresses single diffractive (SD) events by requiring activity on opposite sides of the interaction point. In order to reproduce the results of the analysis, trigger efficiencies for different types of events must be taken into account.
Inelastic events can be either diffractive or non-diffractive (ND).
Since models for diffractive events, {\it e.g.}\ as implemented in {\tt Pythia~8}, produce orders of magnitude fewer antideuterons (per event) than non-diffractive events, we make the approximation that only ND events will contribute to the antideuteron spectrum.
We thus generate pure non-diffractive Monte Carlo events, and re-scale the result according to the fraction of triggered events that are non-diffractive, so that in this approximation predictions for the measured per-event spectrum can be found as 
\beq \label{eq:NDcorrection}
\left. \frac{1}{2\pi N_{\rm ev}} \frac{d^2 N_d}{dp_T dy} \right|_{\rm trig} \simeq f_{\rm ND, trig} \left. \frac{1}{2\pi N_{\rm ev}} \frac{d^2 N_d}{dp_T dy} \right|_{\rm ND}.
\eeq
Here, $N_{\rm ev}$ is the total number of recorded/simulated events and $N_d$ the number of these events with antideuterons. The fraction of triggered events that are non-diffractive is given by
\beq \label{eq:NDtrigger}
f_{\rm ND, trig} \equiv \frac{N_{\rm ND, trig}}{N_{\rm trig}} = \frac{\epsilon_{\rm ND} N_{\rm ND}}{\sum_i \epsilon_i N_i} = \frac{\epsilon_{\rm ND} f_{\rm ND}}{\sum_i  \epsilon_i f_i},
\eeq
where $\epsilon_i$, $N_i$ and $f_i$, respectively, are the trigger efficiency, event count and fraction of the total number of events that are of process type $i$. Event counts and event fractions with the `trig' subscript are events after trigger; the others are before trigger. 
The sum in the denominator is over all inelastic processes: single-, double-, central- (if applicable) and non-diffractive events. 

Trigger efficiencies for the different processes at 0.9, 2.36 and 7~TeV have been estimated using {\tt Pythia~6} and {\tt PHOJET} in Ref.~\cite{etheses2961}.
Using the trigger efficiencies and event fractions for the V0AND trigger from Tables 5.2--5.8 in Ref.~\cite{etheses2961}, we calculate the estimated fraction of triggered events that are non-diffractive according to Eq.~\eqref{eq:NDtrigger}.
The results are listed in Tab.~\ref{tab:ALICEfractions}. 
In our calculations of the ALICE (anti)deuteron spectra, we use the average value of the two Monte Carlo estimates.
For the 2.76~TeV antideuteron events, we use the event fractions calculated for 2.36~TeV as an estimate.

\begin{table}[h!]
\begin{tabular}{l c c c c}
	Energy \slash $f_{\rm ND, trig}$		&\quad&	{\tt Pythia~6} \quad & {\tt PHOJET} \quad	& Average\\ 
    \hline
	0.9~TeV		&&	0.837	&	0.856	&	0.847	\\
	2.36~TeV	&&	0.832	&	0.875	&	0.854	\\
	7~TeV		&&	0.831	&	0.891	&	0.861	\\
\end{tabular}
\caption{Estimated fraction of minimum bias events that pass the ALICE V0AND trigger that are non-diffractive.}
\label{tab:ALICEfractions}
\end{table}

\begin{figure*}
\includegraphics[width=0.45\textwidth]{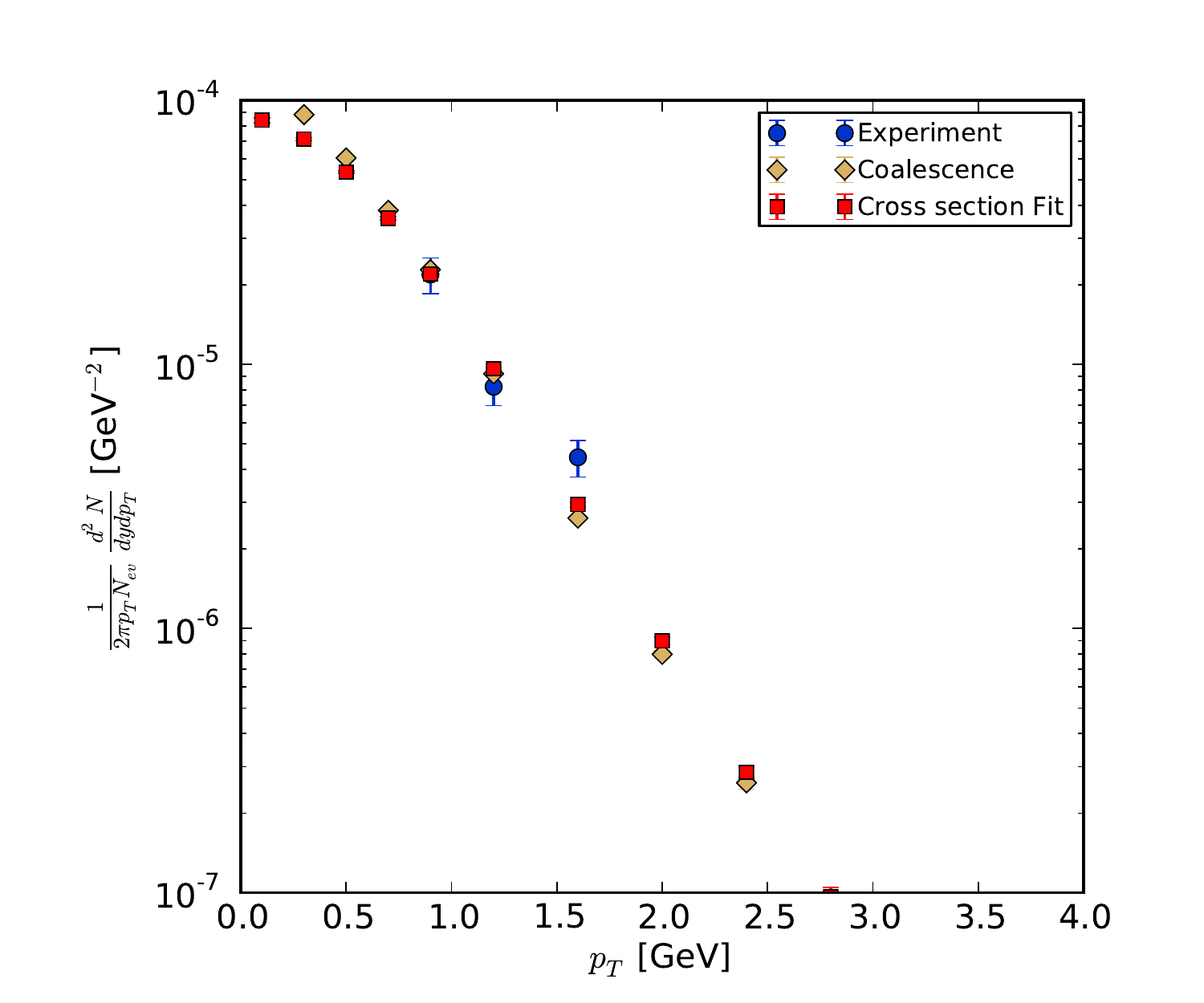}
\includegraphics[width=0.45\textwidth]{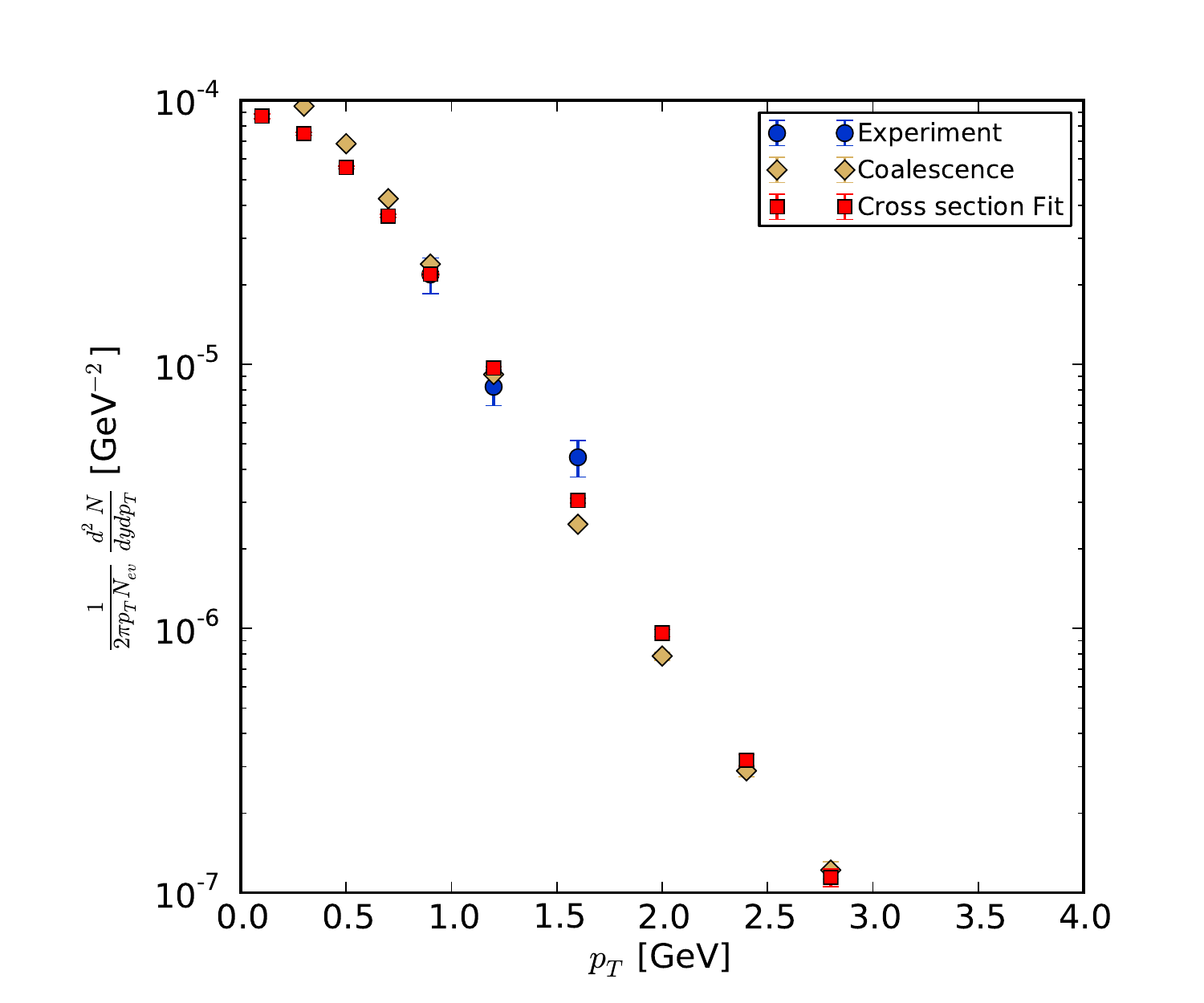}
\includegraphics[width=0.45\textwidth]{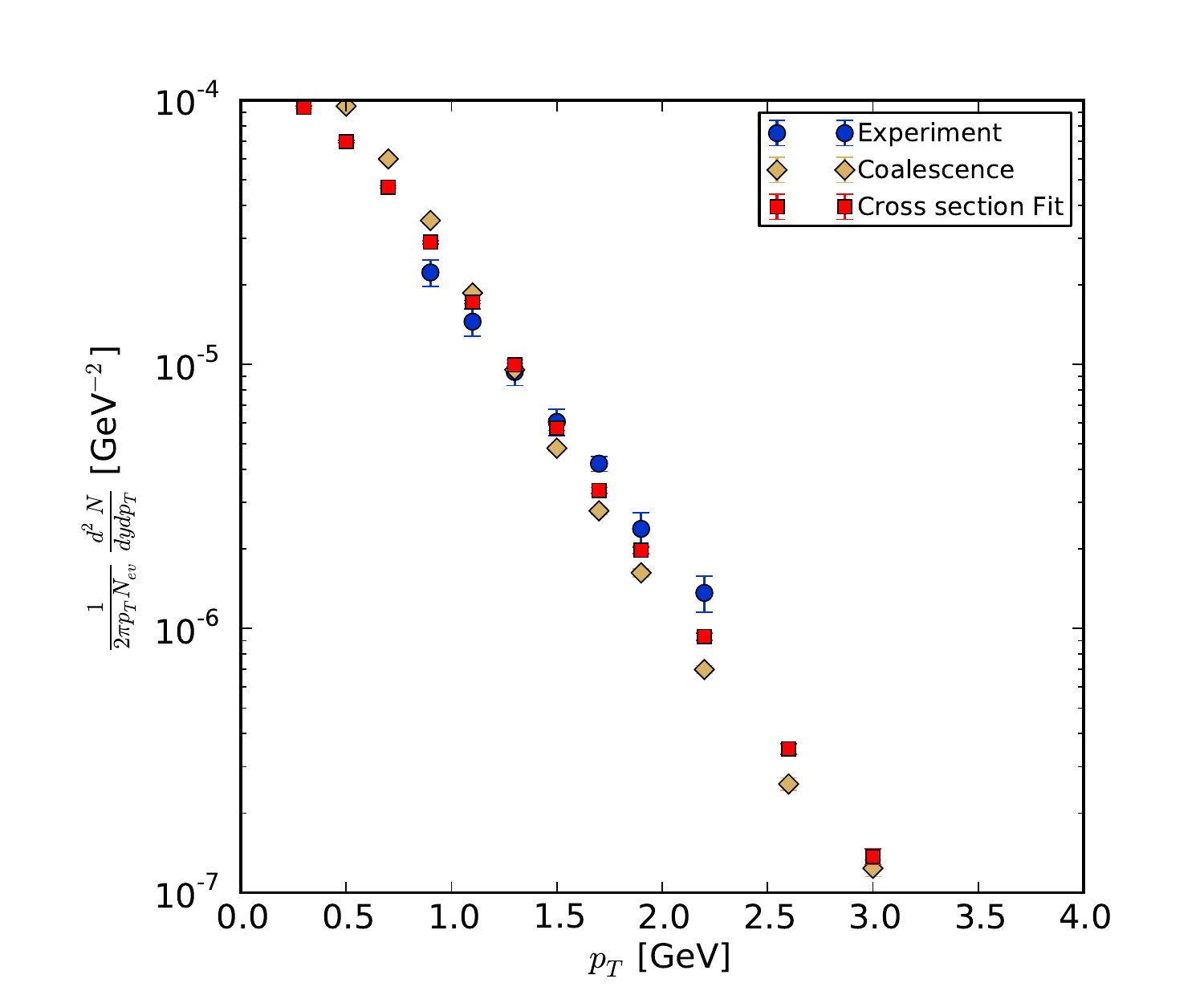}
\includegraphics[width=0.45\textwidth]{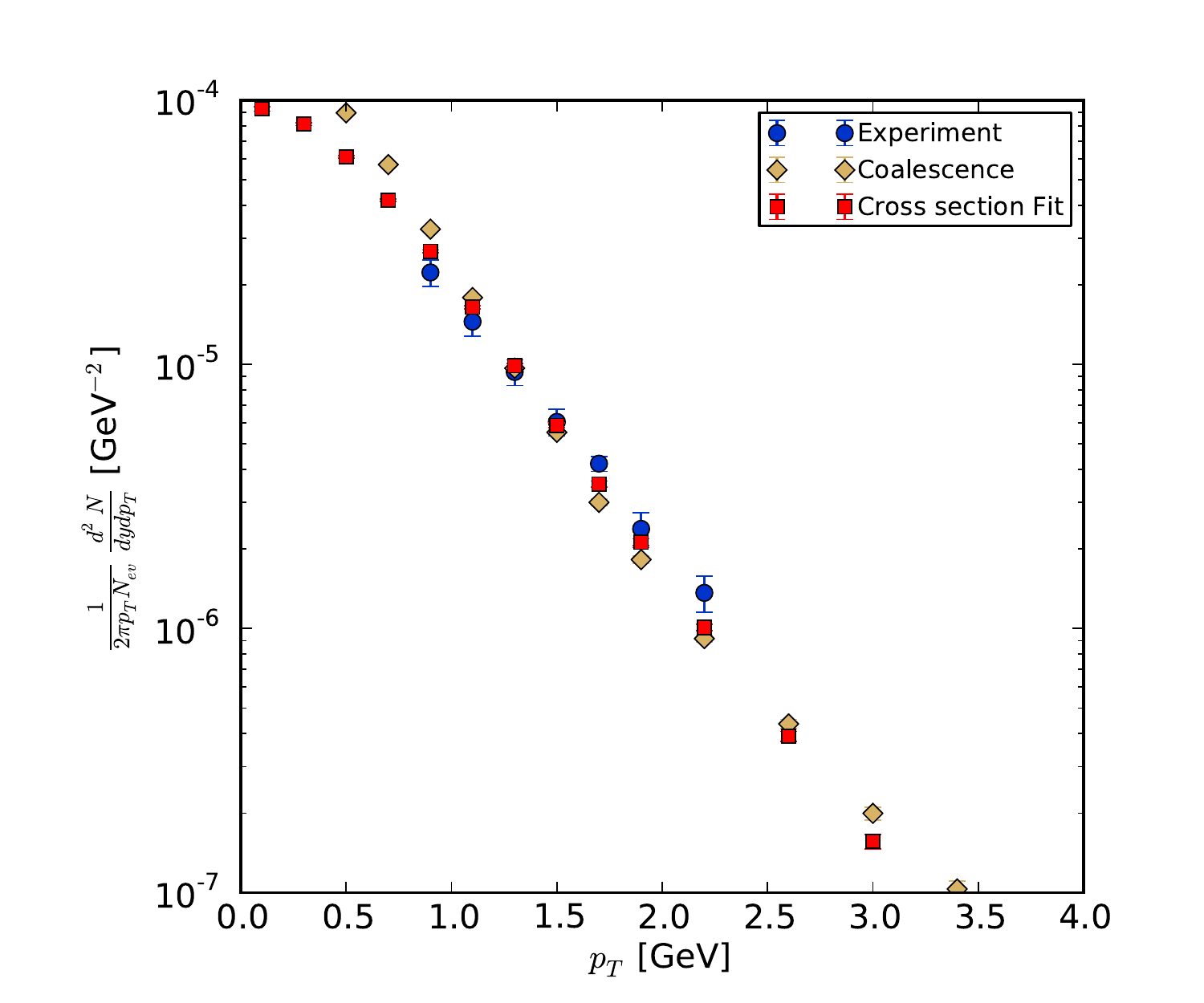}
\includegraphics[width=0.45\textwidth]{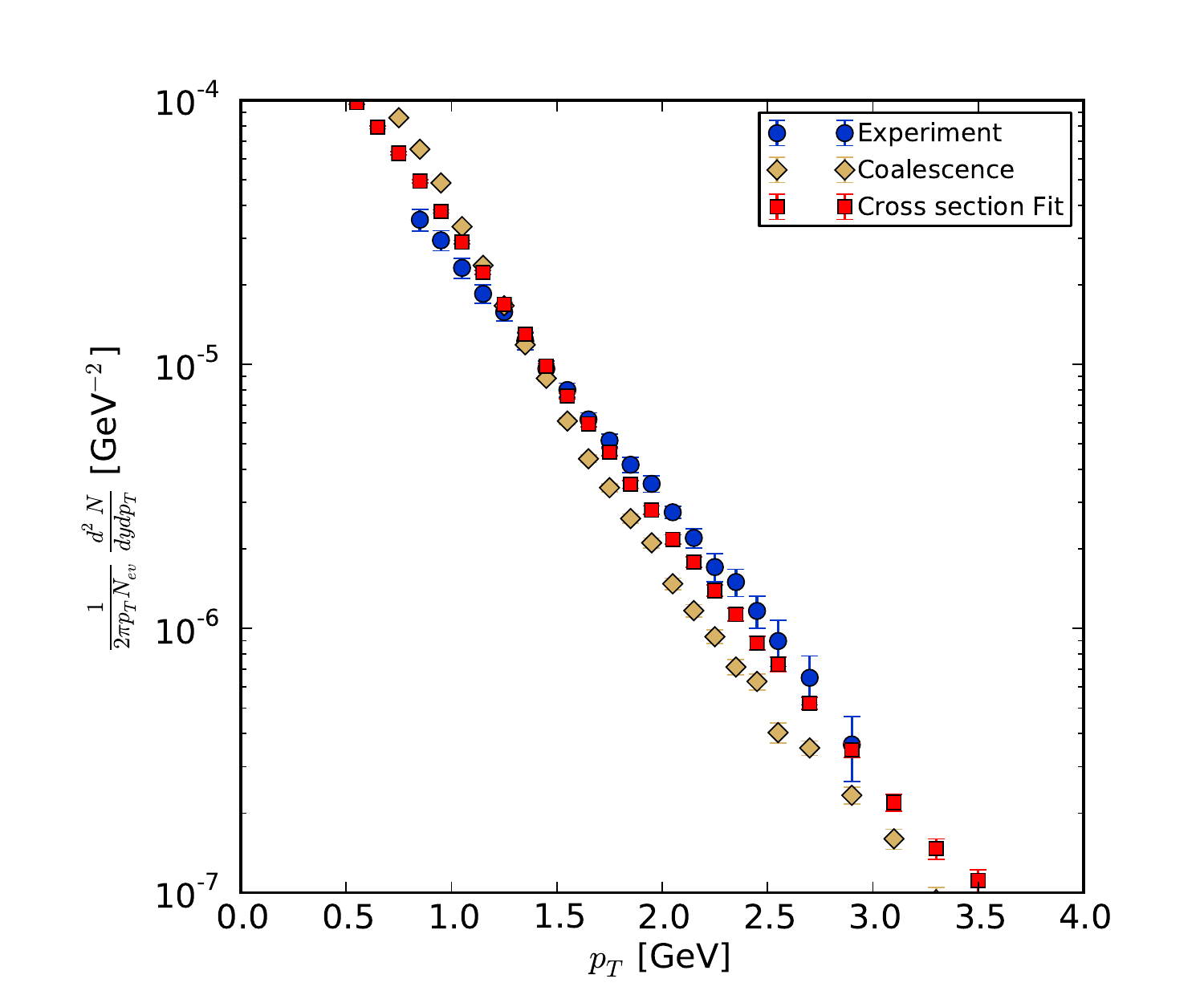}
\includegraphics[width=0.45\textwidth]{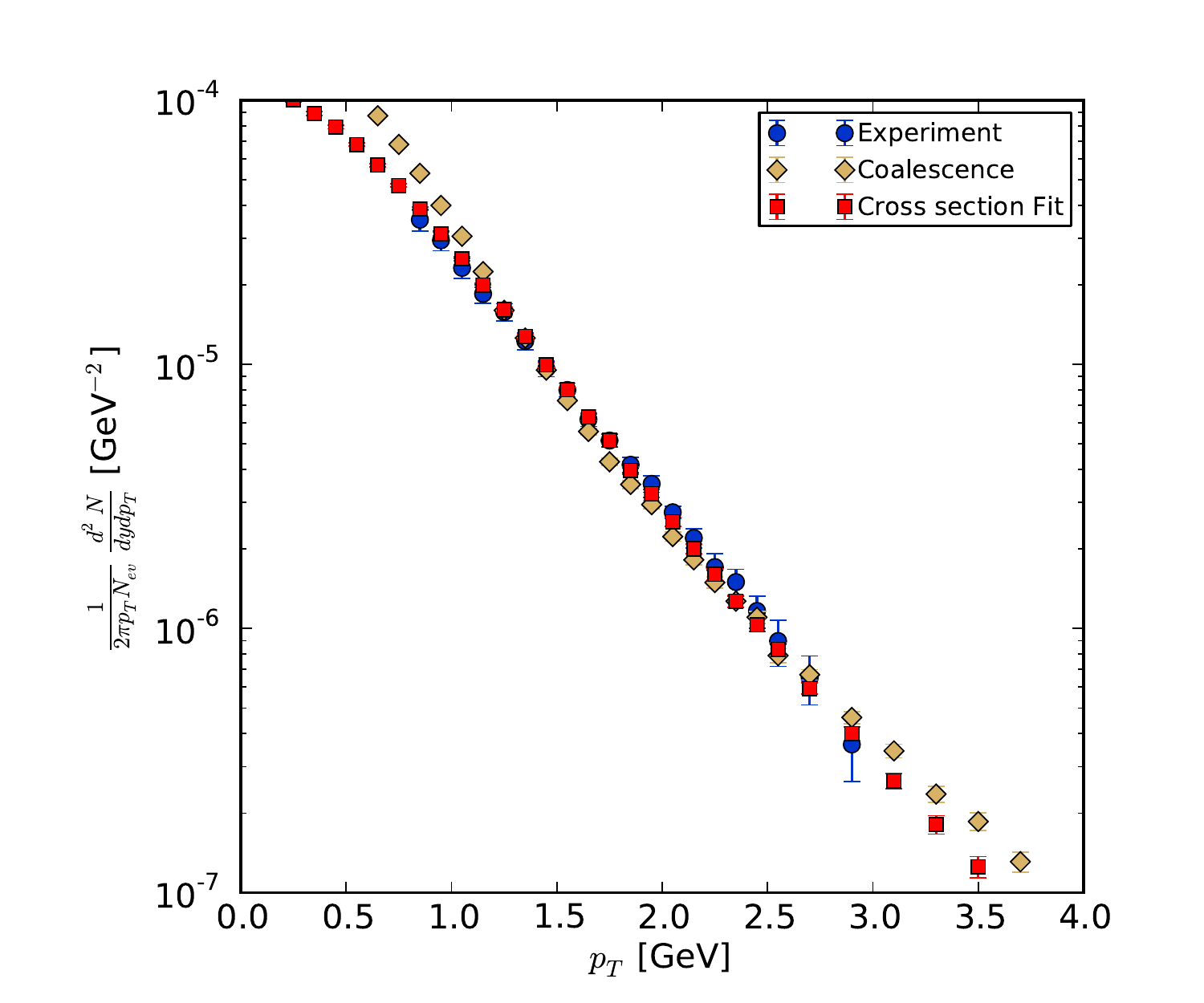}
\caption{Antideuteron spectra at ALICE for the best fit values of $p_0$ and $\sigma_0$ given in Tables~\ref{tab:bestFitHpp} and~\ref{tab:bestFitPy8}. Top: 0.9~TeV, middle: 2.76~TeV, bottom: 7~TeV. Left: {\tt Herwig++}, right: {\tt Pythia~8}.}
\label{fig:ALICE}
\end{figure*}

In Fig.~\ref{fig:ALICE} we show the best fits of both the coalescence model and our cross section based model to the ALICE antideuteron data at all three energies using {\tt Herwig++}  and {\tt Pythia~8}. The fits are done individually at each energy, and the resulting fit values can be found in Tables~\ref{tab:bestFitHpp} and~\ref{tab:bestFitPy8}.
For {\tt Herwig++}, the slope of the spectrum in the coalescence model is quite different from the slope in the experimental data, leading to very bad fits for the large 2.76 and 7~TeV data-sets. 
The cross section model gives a slope that is much closer to the experimental result, however, the fit at 7~TeV is  still quite poor for {\tt Herwig++}. 
For {\tt Pythia~8}, both models give better fits. The shape of the spectrum in the coalescence model still does not match the experimental data. The cross section based model reproduces the shape of the spectrum far better, and gives individual fits with $\chi^2$-values that are consistent with the data, although visually a systematic overshoot at low $p_T$, and undershoot at high $p_T$ seems to be present. This may indicate that ALICE error estimates are somewhat large.

\begin{figure}
\includegraphics[width=0.45\textwidth]{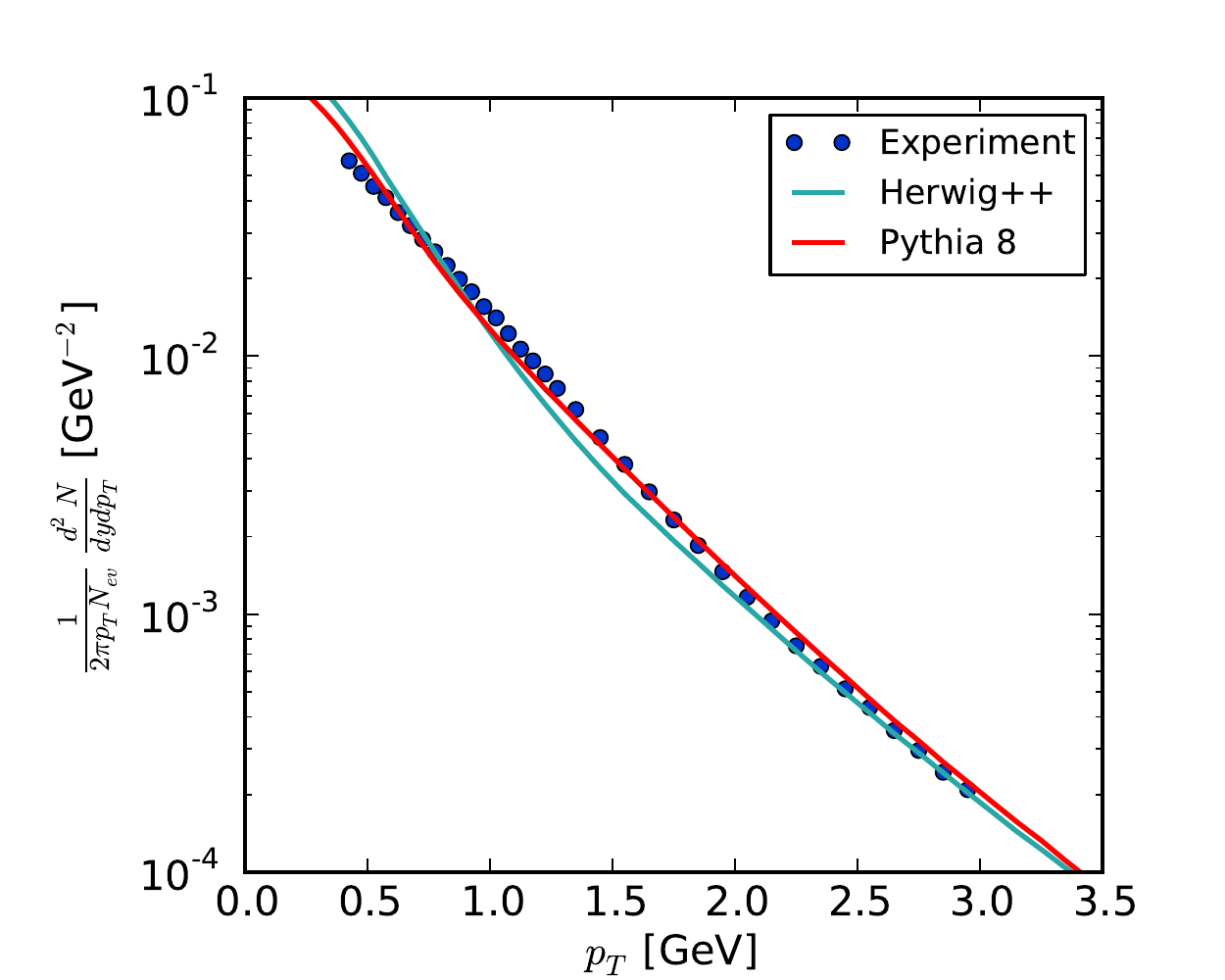}
\caption{Antiproton spectra from ALICE at 7 TeV compared to Monte Carlo models.}
\label{fig:ALICEap}
\end{figure}

In Fig.~\ref{fig:ALICEap} we show the antiproton data from ALICE at 7 TeV~\cite{Serradilla:2013yda} compared to the spectra generated by {\tt Herwig++} and {\tt Pythia~8}. We observe that there are small but systematic differences, most significant for {\tt Herwig++}. In terms of $p_T$-values, these match roughly the intervals where the fits to the antideuteron data are poorest.
Since the antiproton (and antineutron) spectra are the basis for the antideuteron spectrum, it is unreasonable to expect that we can reproduce the antideuteron better than the progenitors. It is interesting to speculate if one could retune the generators so that they would better reproduce the antiproton data, similar to~\cite{Dal:2014nda}, and how significant the improvement would be for the antideuterons. However, this is outside the scope of the present paper, which focuses on the antideuteron production model itself.

As previously mentioned, while the coalescence model can give good fits to individual experiments, the best fit value of the coalescence momentum $p_0$ varies strongly between different experiments. This has lead to speculation that there should be some dependence on the COM energy of the process in the antideuteron production model. 
In {\tt Herwig++}, (see Table~\ref{tab:bestFitHpp})  we see a significant energy dependence in the best fit values of $p_0$ and $1/\sigma_0$, with higher energies requiring lower values. 
In the coalescence model, the spectrum scales roughly as $p_0^3$, while in the cross section model it scales as $1/\sigma_0$.
Based on this, the energy dependence is significantly stronger in the coalescence model than in the cross section model.
In contrast, in {\tt Pythia~8}, we see a much weaker energy dependence in both models. Here the energy dependence is somewhat more significant in the cross section based model. While we see some sign of energy dependence in both deuteron and antideuteron production, it is still unclear if any of this dependence can be attributed to the production models themselves, or if this is entirely an effect of the Monte Carlos used.

\subsection{BABAR}
Antideuteron production in $\Upsilon(1S,2S,3S)$ decays and non-resonant $e^+e^-\rightarrow q\bar q$ processes at $\sqrt{s} = 10.58$~GeV has been measured by the BABAR Collaboration~\cite{Lees:2014iub}.
The latter is of particular interest for DM studies, as it resembles the primary annihilation process in many DM scenarios with a two-particle colorless (electro)weak initial state. It can also be directly compared to LEP results (see below) at $\sqrt{s}=91.2$ GeV.

\begin{figure*}
\includegraphics[width=0.35\textwidth]{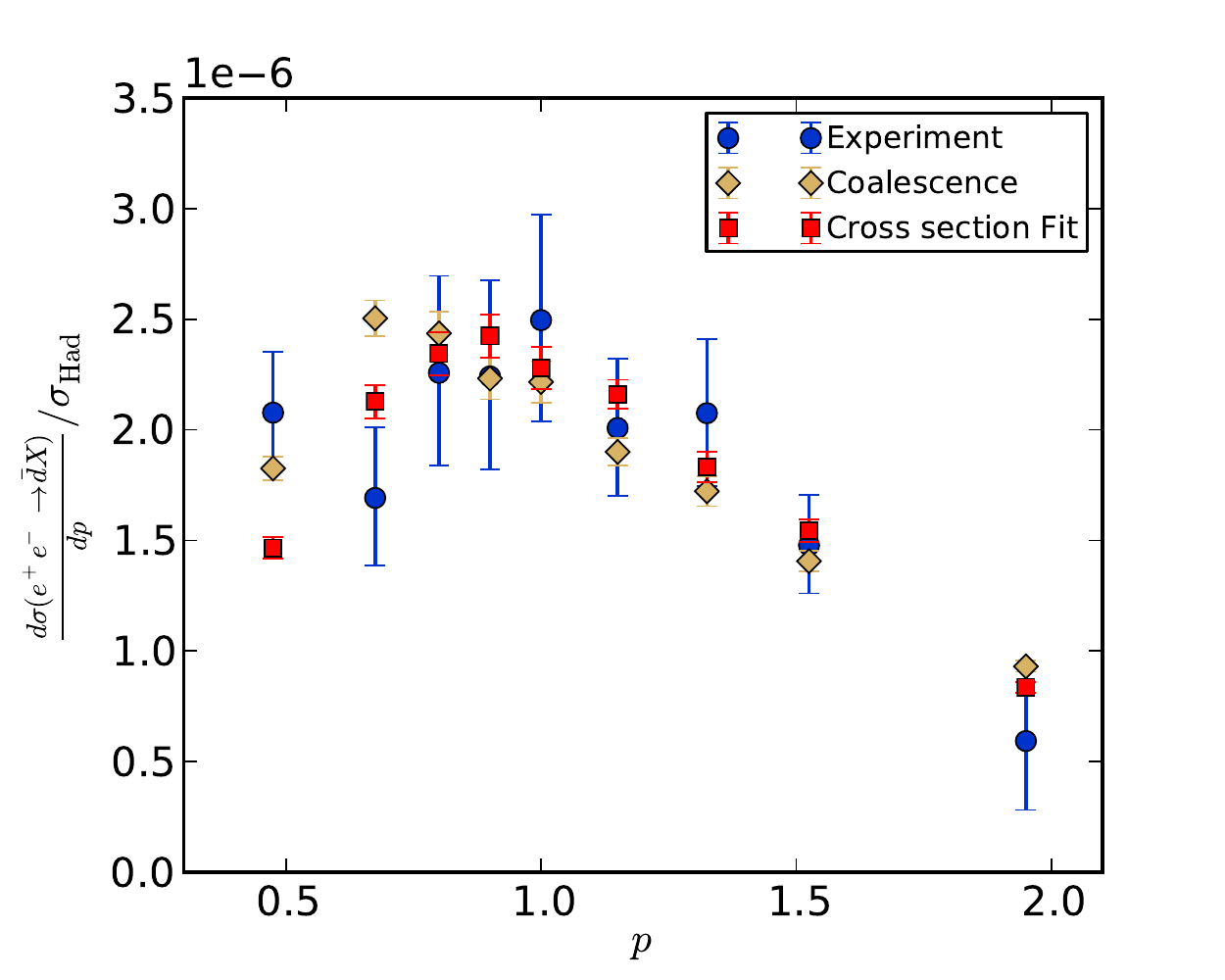}
\includegraphics[width=0.35\textwidth]{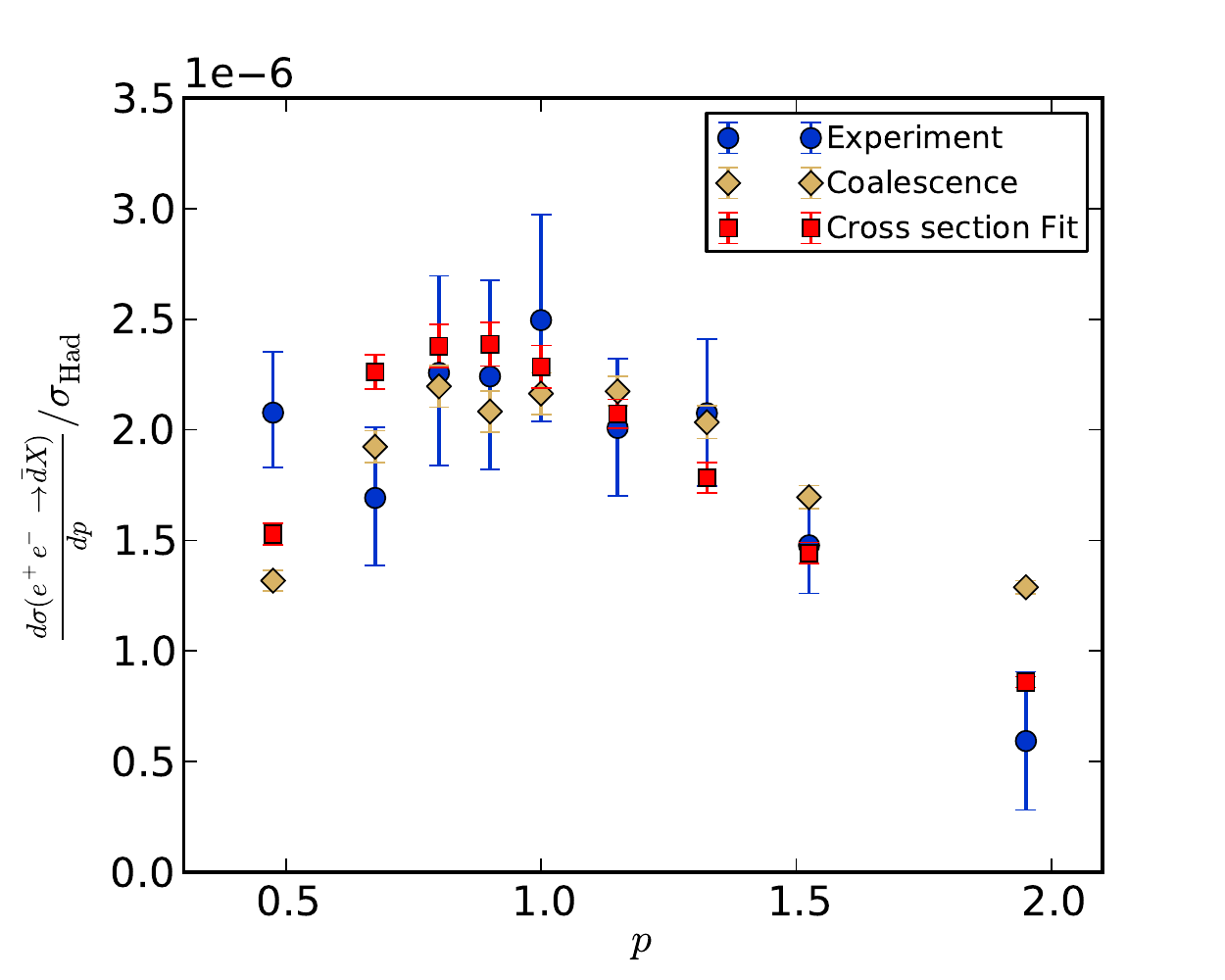}
\caption{Antideuteron spectra at BABAR for the best fit values of $p_0$ and $\sigma_0$ given in Tables~\ref{tab:bestFitHpp} and~\ref{tab:bestFitPy8}. Left: {\tt Herwig++}, right: {\tt Pythia~8}.}
\label{fig:BABAR}
\end{figure*}

The cross section based model gives better fits to the data than the coalescence model in both Monte Carlos, see Fig.~\ref{fig:BABAR} and Tables~\ref{tab:bestFitHpp} and~\ref{tab:bestFitPy8}. In {\tt Herwig++}, the difference is rather small as the spectra from the two models differ mainly at low energies, where the experimental data fluctuates with large errors in the two lowest energy bins. With {\tt Pythia~8}, the cross section based model gives a notably better fit, as it gives a better description of the high energy data.

In {\tt Pythia~8}, we find the best fit parameters to the continuum process to be reasonably similar to those obtained from $\Upsilon(1S)$ decays at CLEO (see below). This is not entirely unexpected, given that the two processes have very similar energies and a colorless initial state. However, in {\tt Herwig++},  we find the best fit values of BABAR to lie far below the best fit values from CLEO. This difference in best fit values correspond to a factor $\sim3$ in antideuteron production for the coalescence model, while the difference is around 50\% in the cross section model.
The explanation for this that is nearest at hand, is that while the processes have similar energies, $\Upsilon(1S)$ decays into gluonic final states rather than quark final states, and that this brings into play differences in the two Monte Carlo generators. We have checked that the two Monte Carlos produce similar antinucleon multiplicities in each of the two experiments, 
which seems to imply that there is a strong process dependence in the two-(anti)baryon correlations from the {\tt Herwig++} cluster hadronization model at these energies. 

Compared to the ALICE data from proton-proton collisions the fitted values of $p_0$ and $1/\sigma_0$ are significantly smaller for both generators. This trend continues below for the other $e^+e^-$-experiments. It is difficult to determine if this is somehow a process dependence that should be incorporated into a more complete model, or alternatively an energy dependence, as the $e^+e^-$-experiments are typically at much lower energies. The latter has some support in the trend for better agreement at larger values of COM-energy seen for the LEP results.

\subsection{CERN ISR}
The antideuteron spectrum in $pp$-collisions at $\sqrt{s}=53$~GeV were measured at the CERN Intersecting Storage Rings (ISR) at $\theta_{\rm lab} = 90\degree$~\cite{Henning:1977mt} and $\theta_{\rm lab} = 62.5\degree$~\cite{Alper:1973my}.

In our Monte Carlo analysis, we generate minimum bias events.
As discussed in the ALICE analysis section, antideuterons are hardly produced in diffractive events, and we therefore generate purely non-diffractive events, and use the corresponding non-diffractive Monte Carlo cross sections in calculating the invariant cross section $Ed^3\sigma/dp^3$.\footnote{In our previous work~\cite{Dal:2014nda}, we used the larger experimentally measured total inelastic cross section, leading to an overestimation of the antideuteron yield.}
We note that the cross section from {\tt Pythia~8} is a factor $\sim 22\%$ larger than the cross section from {\tt Herwig++}.
Since the yield is absolute, and not per event, this leads to an artificial difference in the antideuteron yield that is not related to the event generation itself, and the difference in cross section should therefore be kept in mind when comparing best fit parameters of the two antideuteron production models.

\begin{figure*}
\includegraphics[width=0.35\textwidth]{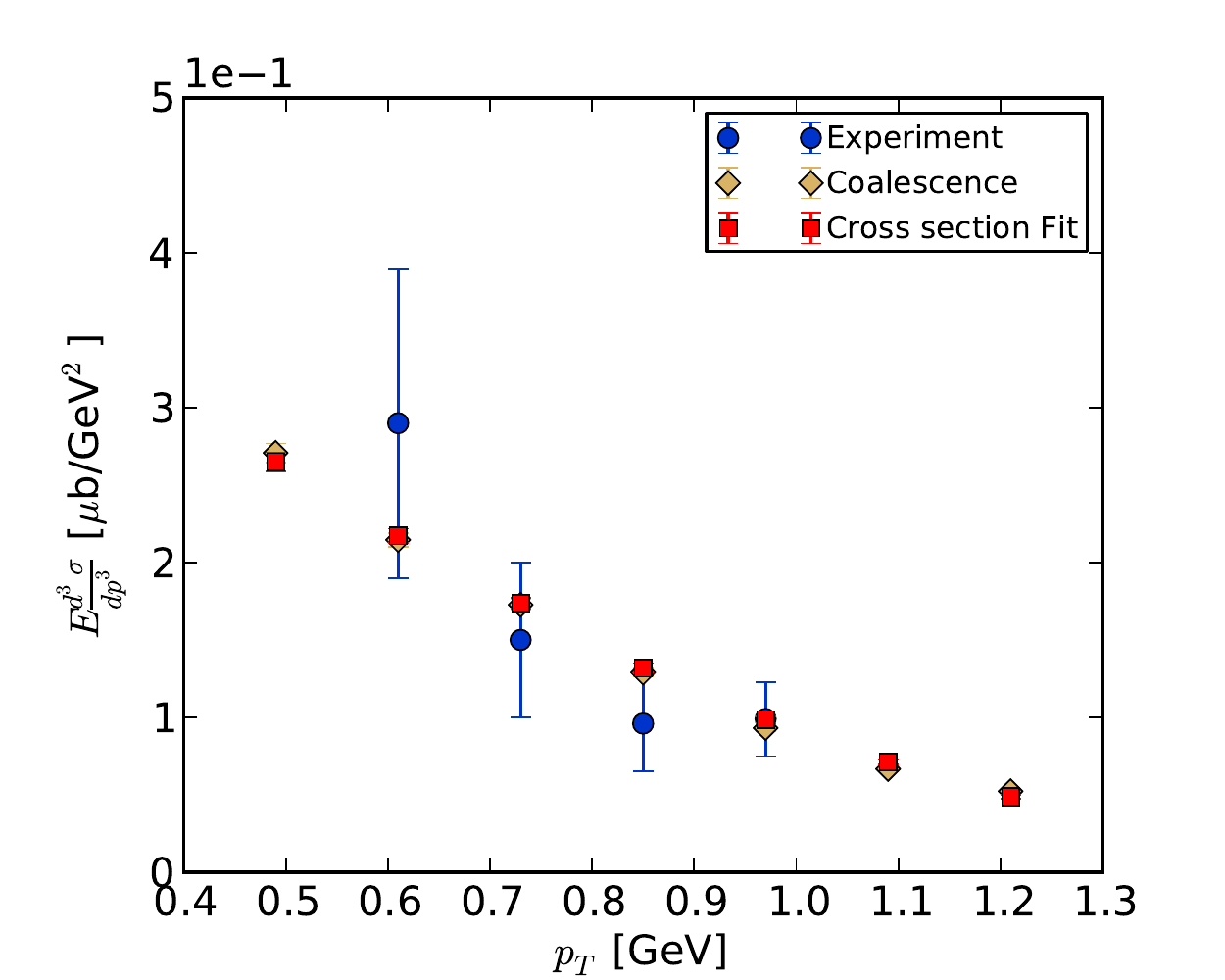}
\includegraphics[width=0.35\textwidth]{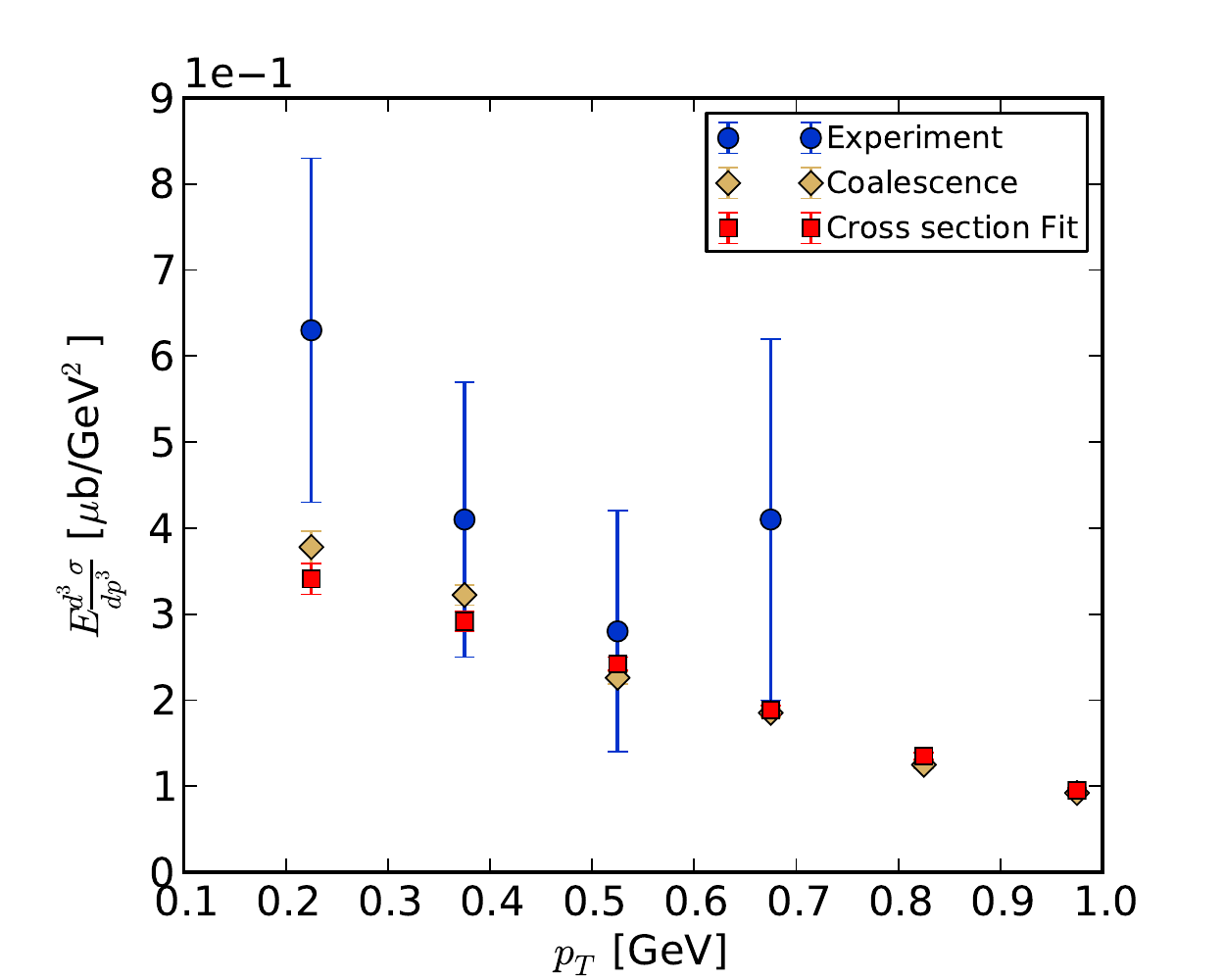}
\caption{Antideuteron spectra at ISR, generated using {\tt Herwig++} with the best fit values of $p_0$ and $\sigma_0$ given in Tab.~\ref{tab:bestFitHpp}. 
		 Left: $\theta_{\rm lab} = 90\degree$, right: $\theta_{\rm lab} = 62.5\degree$.}
\label{fig:ISR_Hpp}
\end{figure*}

\begin{figure*}
\includegraphics[width=0.35\textwidth]{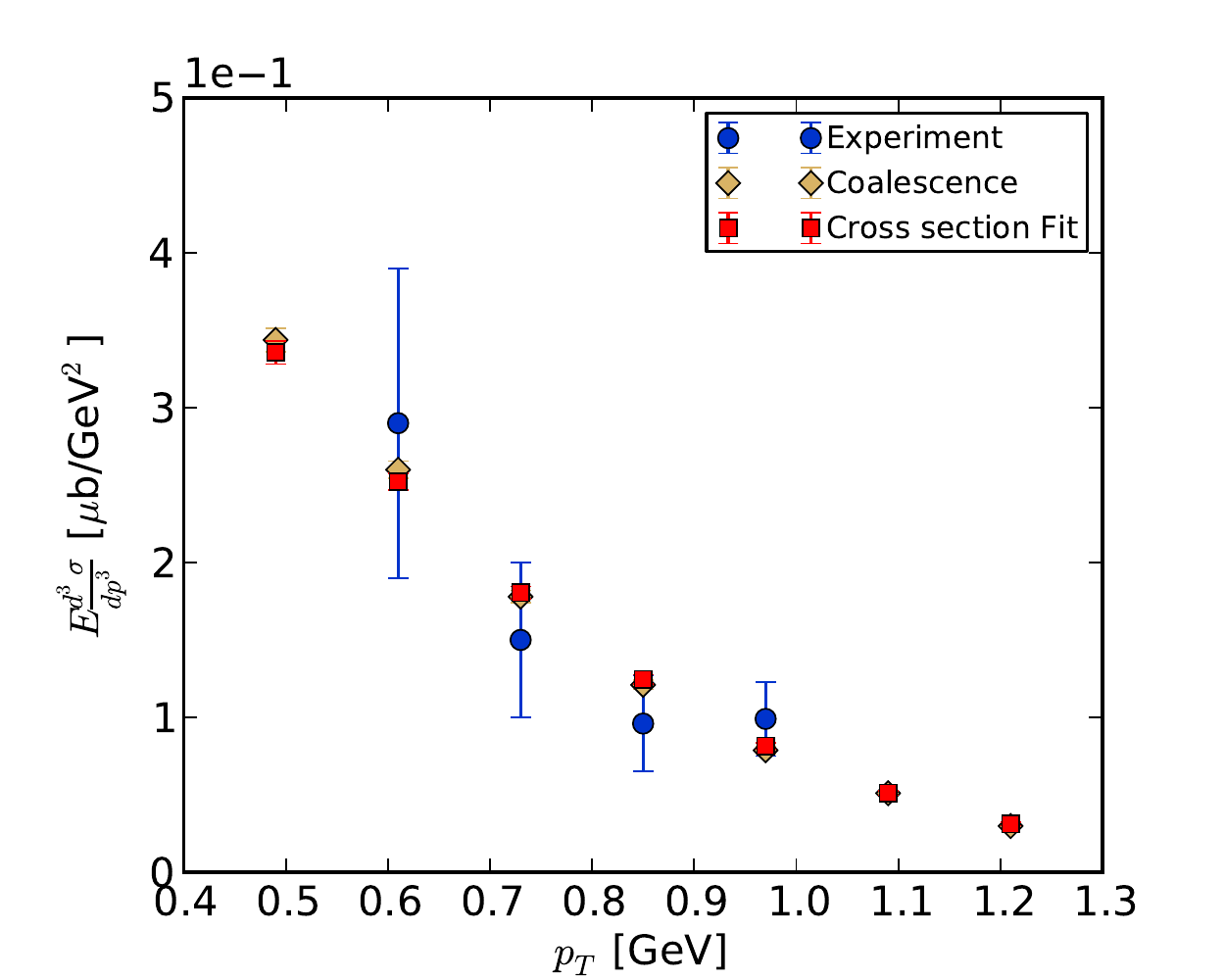}
\includegraphics[width=0.35\textwidth]{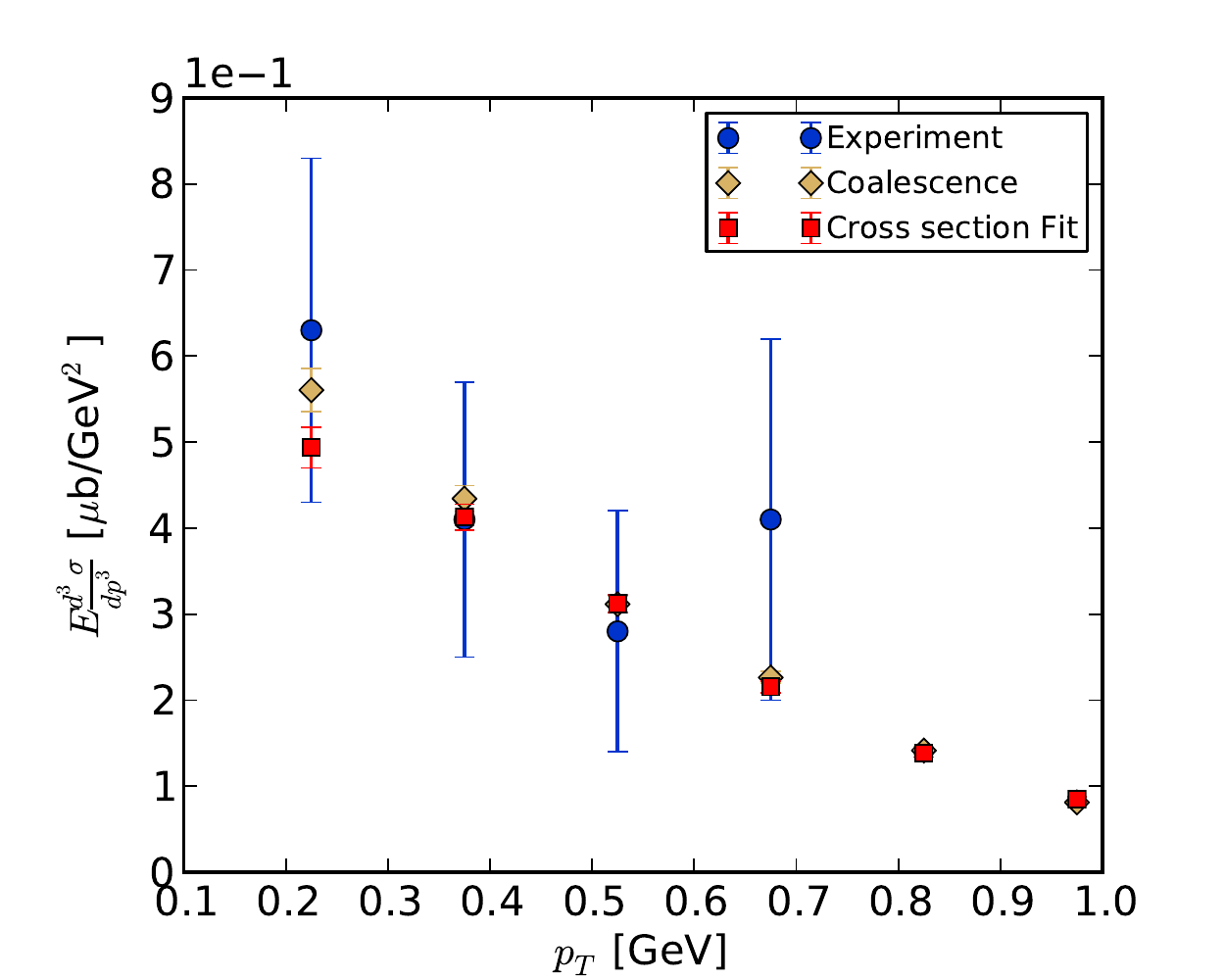}
\caption{Antideuteron spectra at ISR, generated using {\tt Pythia~8} with the best fit values of $p_0$ and $\sigma_0$ given in Tab.~\ref{tab:bestFitPy8}. 
		 Left: $\theta_{\rm lab} = 90\degree$, right: $\theta_{\rm lab} = 62.5\degree$.}
\label{fig:ISR_Py}
\end{figure*}

We perform a combined fit to the two datasets, and the spectra are plotted using the combined best fit values of $p_0$ and $\sigma_0$ in Figs.~\ref{fig:ISR_Hpp} and~\ref{fig:ISR_Py}.
The two antideuteron formation models produce quite similar results in both Monte Carlos, and due to the large experimental errors, the differences in $\chi^2$ are small --- the coalescence model giving a slightly better fit. 
We find the two Monte Carlos to give wildly different best fit values of $p_0$ and $\sigma_0$:
{\tt Herwig++} gives unusually large best fit values, whereas {\tt Pythia~8} gives moderately low values compared to ALICE.
The difference in cross section between the Monte Carlos constitutes only a small part of this difference.
We have checked that {\tt Pythia~8} produces a 37\% higher multiplicity of antinucleons per event at this energy, and this difference is likely responsible for a sizeable part of the discrepancy.

The best fit values of $p_0$ and $\sigma_0$ differ significantly from the values found for the ALICE measurements in both Monte Carlos. In {\tt Herwig++}, we see a continuation of the trend of increasing values with decreasing COM energies, and this may be another indication of an energy dependence stemming from the cluster hadronization model.
While we saw a similar tendency in the {\tt Pythia~8} ALICE results, the ISR results do not support the hypothesis of a systematic COM energy dependence in this Monte Carlo.

\subsection{CLEO}
\begin{figure*}
\includegraphics[width=0.35\textwidth]{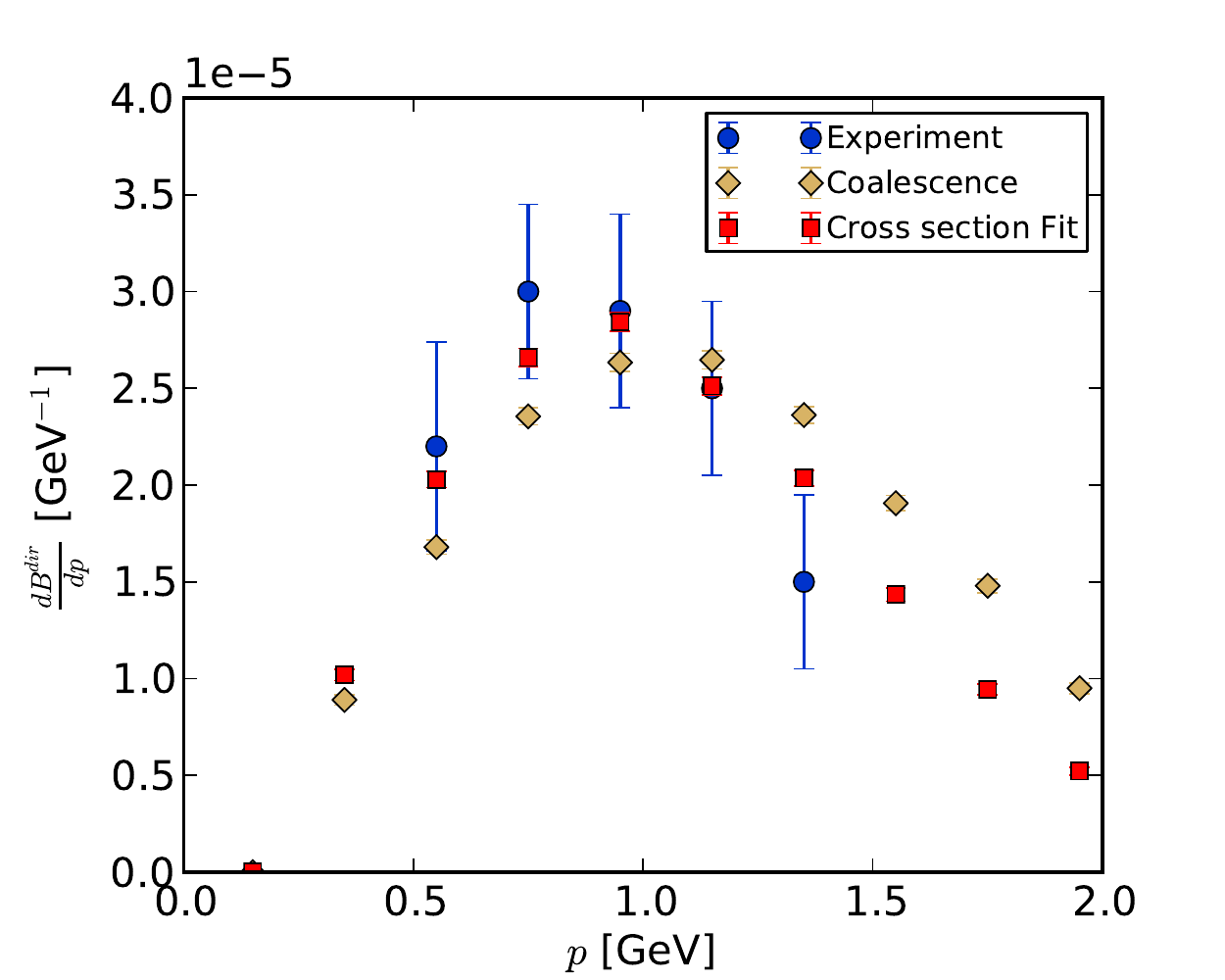}
\includegraphics[width=0.35\textwidth]{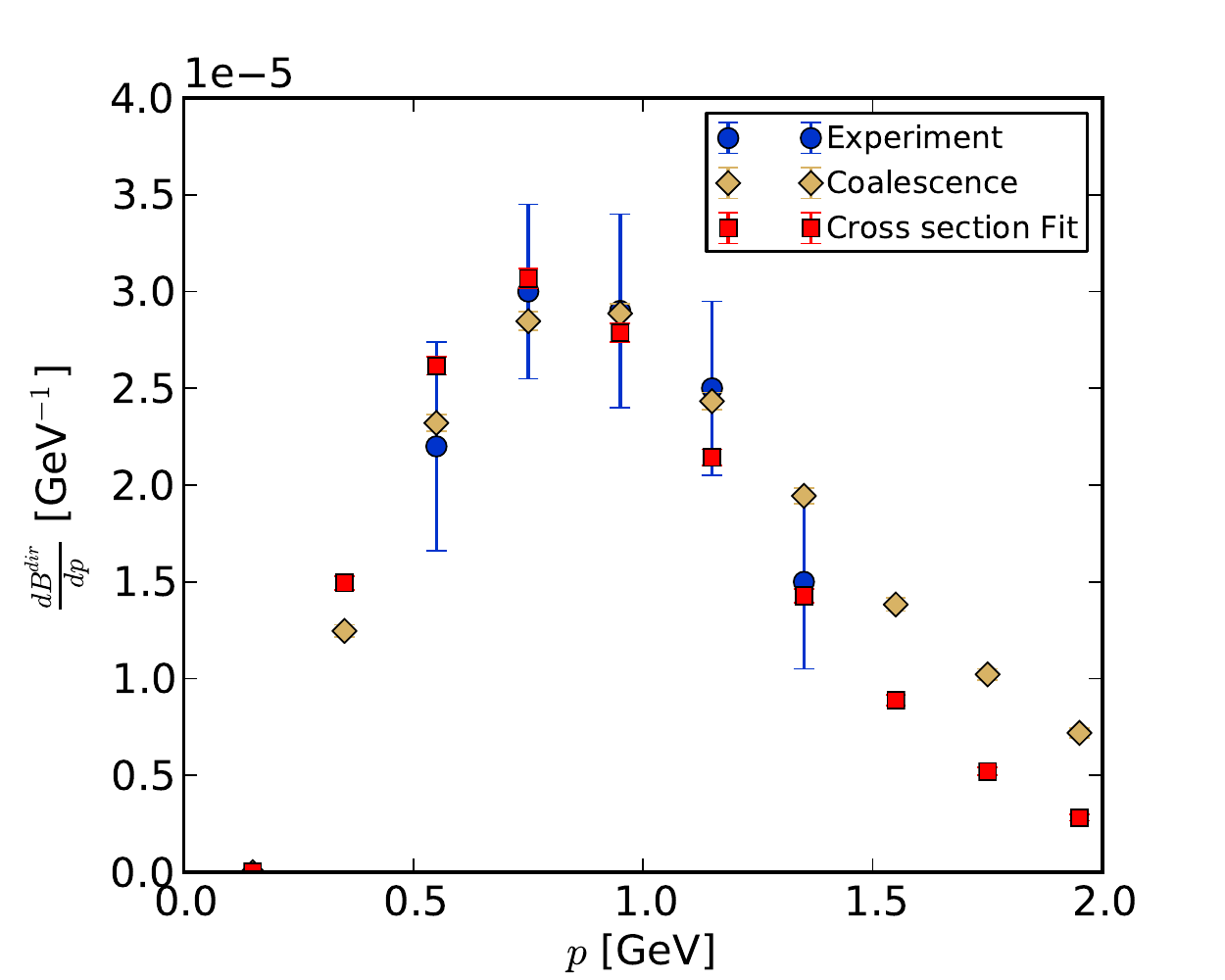}
\caption{Antideuteron spectra at CLEO for the best fit values of $p_0$ and $\sigma_0$ given in Tab.~\ref{tab:bestFitHpp}. Left: {\tt Herwig++}, right: {\tt Pythia~8}.}
\label{fig:CLEO}
\end{figure*}

Antideuteron production in $\Upsilon(1S)$ decays has been measured at CLEO~\cite{Asner:2006pw}.
The best fit spectra for the CLEO data are shown in Fig.~\ref{fig:CLEO}. 
{\tt Herwig++} and {\tt Pythia~8} have similar best fit values of $p_0$ in the coalescence model, while the best fit $\sigma_0$ differs quite significantly in the cross section based model.
In {\tt Herwig++}, the cross section based model reproduces the shape of the spectrum significantly better than the coalescence model, and thus gives a better fit. 
In {\tt Pythia~8}, the coalescence model gives a better fit to the low energy data, while the cross section based model gives a better fit to the high energy data. As a result the two models give very similar fits; the coalescence model having a slightly lower $\chi^2$.

\subsection{LEP}
Antideuteron searches in $e^+e^-$-collisions were performed by the ALEPH~\cite{Schael:2006fd} and OPAL~\cite{Akers:1995az} experiments at LEP.
Both collaborations studied antideuteron multiplicities in hadronic events at the Z-resonance.
ALEPH observed $(5.9 \pm 1.8 \pm 0.5)\e{-6}$ antideuterons per hadronic event in the momentum range $0.62 < p < 1.03$~GeV and angular range $|\cos \theta| < 0.95$, where the errors are statistical and systematical, respectively.
In OPAL, however, no antideuteron candidates were observed in the antideuteron momentum range $0.35<p<1.1$~GeV.
In previous works, only the ALEPH result has been used for calibration of the the coalescence momentum, but the negative OPAL result should also be taken into account.
As the expected number of signal events in the two experiments are of the same magnitude, 
the non-observation of antideuterons in OPAL might be an indication that the ALEPH result suffers from an upwards fluctuation.
Performing a combined $\chi^2$ fit of the two experiments will yield a lower best fit coalescence momentum than previous fits based on the ALEPH data alone.

In order to calculate the $\chi^2$ for OPAL, we first estimate the expected number of signal events $s$ at OPAL by
\beq
 s = \epsilon N_{ev} n_{\bar d, \rm MC},
\eeq
where $\epsilon = 0.234$ is the given detection efficiency, $N_{ev} = 1.64\e{6}$ is the number of events in the OPAL analysis, and $n_{\bar d, \rm MC}$ is the Monte Carlo prediction for the number of antideuterons per event.
$\epsilon$ and $n_{\bar d, \rm MC}$ are in reality energy dependent quantities, but only the average value of the detection efficiency is available.
Using the fact that no antideuteron candidates were observed by OPAL, and assuming Poissonian uncertainty $\sigma=\sqrt{s}$ for the expected number of events, the $\chi^2$ is then given by
\beq
  \chi^2_{\rm OPAL} = \frac{(N_{\rm obs} -s)^2}{\sigma^2} = s.
\eeq

\subsection{Combined fits} \label{sec:fitValues}
For the purpose of calculating the cosmic ray antideuteron flux from dark matter, the free parameters $p_0$ and $\sigma_0$ have to be calibrated based on fits to the previously discussed experimental data. This calibration should be done separately for each Monte Carlo, as the best fit values generally differ between Monte Carlos, which should be clear from the above, {\it e.g.}\ due to differences in primary antinucleon spectra and the (anti)nucleon correlations. 
It is also clear that no parameter values exist that give good simultaneous fits to all experiments, and this can be seen quantitatively in Table~\ref{tab:combinedFit}: combined fits to all experiments yield $\chi^2$/d.o.f ranging from 10 to a whopping 54.

\begin{table*}
\begin{tabular}{l l l c c c c c c c c c}
	Monte Carlo		&\quad&	Experiments					& Data points	&\quad& Best fit $p_0$ [MeV]	&\quad& $\chi^2_{p_0}$ &\quad& Best fit $1/\sigma_0$ [$\rm barn^{-1}$]	&\quad& $\chi^2_{\sigma_0}$	\\ 
    \hline
	{\tt Herwig++}	&&	ALICE ($\rm \bar d$), ISR	    &	38		&&		187			&&	646		&&	3.50	&&	196		\\
					&&	BABAR, CLEO, LEP				&	16		&&		96			&&	73.6	&&	0.68	&&	29.2	\\
					&&	All experiments					&	54		&&		123			&&	2859	&&	1.43	&&	2146	\\
	{\tt Pythia~8}	&&	ALICE ($\rm \bar d$), ISR	    &	38		&&		193			&&	255		&&	2.63	&&	58.2	\\
					&&	BABAR, CLEO, LEP				&	16		&&		140			&&	30.5	&&	1.18	&&	16.7	\\
					&&	All experiments					&	54		&&		174			&&	888		&&	2.13	&&	510		\\
\end{tabular}
\caption{Combined best fit parameters for the coalescence model and the cross section model for different selections of experimental data.}
\label{tab:combinedFit}
\end{table*}

In order to get sensible results, it is necessary to restrict the fits to reasonably self-consistent subsets of experiments.  
However, it is not  {\it a priori} clear which experiments should be included in the fits.
In previous work, the choice of $p_0$ for the coalescence model has often been based on a fit to the ALEPH data alone, as LEP events are similar to DM annihilation events in typical DM models. However, as discussed earlier, the OPAL experiment at LEP did not observe any antideuterons in a similar range of energies. In fact, even the ALEPH data alone does not constitute more than a 3$\sigma$ observation of antideuterons. The problem of relying on a single data point for calibration has also been discussed extensively in the past, {\it e.g.}\ see \cite{Dal:2012my}.

We will here divide the experiments into two groups: experiments with colored initial states (ALICE, ISR), and experiments with colorless initial states (BABAR, CLEO, LEP), and consider separate fits to these two sets.
While the antideuteron formation process is in both models assumed to be agnostic to the nature of the hard process, it is not unlikely that the outcomes of the hadronization models of the Monte Carlos are sensitive to differences in the underlying physics between these two classes of processes, and thus originate a difference in the fitted value.
As dark matter annihilations have colorless initial states, the colorless set is likely the most relevant for calculating the cosmic ray antideuteron flux from dark matter. 

Best fit values for the two sets of experiments can be seen in Table~\ref{tab:combinedFit}. 
In all cases, the cross section based model gives a considerably better combined fit than the coalescence model.
In {\tt Herwig++}, the fits are still rather bad for both datasets with either model. This is not entirely unexpected;
in the colorless set, the BABAR data prefers much lower values of the free parameters than the other experiments, thus giving a bad simultaneous fit. In the set with colored initial states, the poor individual fits of {\tt Herwig++} to the ALICE data alone are enough to give a bad combined fit, and the large spread in the individual best fit parameters further worsens the result.

In {\tt Pythia~8}, the coalescence model gives relatively poor fits to both sets. 
The cross section based model, on the other hand, gives a good fit to the set with colorless initial states with $\chi^2=16.7$ for 15 degrees of freedom (d.o.f), and gives a decent fit with $\chi^2=58.2$ for 37 d.o.f to the set with colored initial states.

\section{Dark matter spectra}
\label{sec:DM}
We will here compare the antideuteron spectra at Earth coming from a generic dark matter candidate annihilating into $b\bar b$ and $W^+W^-$ in the coalescence and cross section based models, in order to see the impact of the new model on the, in principle, measurable spectrum.  We will be using the best fit values of $p_0$ and $\sigma_0$ discussed in Sec.~\ref{sec:fitValues}, and we will consider dark matter candidates with masses of 100, 500 and 1000~GeV.

\subsection{Propagation of antideuterons}
\label{sec:prop}
Antideuterons, being charged particles, do not propagate through our galaxy in straight lines, but are deflected in the turbulent Galactic magnetic fields.
This leads to a random walk behaviour, which can be well described using a diffusion model.
The most commonly used model is the so-called two-zone diffusion model -- a cylindrical model consisting of a magnetic halo region of radius $R=20$~kpc and half-height $L$, where charged particles diffuse freely; 
and a thin gaseous disk of the same radius and a half-height of $h = 100$~pc, where scattering and annihilation on interstellar matter can additionally take place. 
While $R$ and $h$ are set by the size of the observed Galactic disk, the half-height of the magnetic halo, $L$, is a free parameter.

For antideuterons, energy redistribution terms and non-annihilating inelastic scattering only constitute minor corrections, and are typically neglected, as they will be here.
Under the assumption of steady state conditions, the diffusion equation describing this model is then given by
\beq \label{eq:Diffusion}
 -D(T) \nabla^2 f  +  \frac{\partial}{\partial z} ( {\rm sign}(z) f V_c) = 
 Q - 2 h \delta (z) \Gamma_{\rm ann}(T) f \,,
\eeq
where $f(\vec{x},T)=dN_{\bar{d}}/dT$ is the number density of antideuterons 
per unit kinetic energy $T$, $D(T)= D_0 \beta \mathcal{R}^\delta$ is the 
(spatial) diffusion coefficient, $Q$ is the source term from dark matter annihilations, $V_c$ is the velocity of a convective wind perpendicular to the Galactic disk, $z$ is the vertical coordinate, $\beta=v/c$ is the antideuteron velocity, and
$\mathcal{R}$ is the antideuteron rigidity in units of GV.  $\delta$, $D_0$, and $V_c$ are here free parameters of the model.

The annihilation rate, $\Gamma_{\rm ann} $, of antideuterons on interstellar gas in the Galactic disk is given by
\beq
\Gamma_{\rm ann}(T)  = (n_{H} + 4^{\frac{2}{3}} n_{\rm {He}}) 
                  v \sigma^{\rm {ann}}_{\bar{d}p}(T) ,
\eeq
where $n_H \approx 1 \unit{cm^{-3}}$ and $n_{He} \approx 0.07 n_H$ are the respective number densities of hydrogen and helium nuclei in the disk.
The factor $4^{\frac{2}{3}}$ here accounts for the difference in annihilation cross section between H and He, under the assumption of simple geometrical scaling.
We estimate the annihilation cross section using
\beq
\sigma^{\rm {ann}}_{\bar{d}p}(T) = \sigma^{\rm {tot}}_{\bar{d}p}(T)-\sigma^{\rm {el}}_{\bar{d}p}(T)-\sigma^{\rm {inel,non-ann}}_{\bar{d}p}(T),
\eeq
where $\sigma^{\rm {inel,non-ann}}_{\bar{d}p} = \sigma (\bar d p \rightarrow  \bar d X)$ is the component of the inelastic cross section where the antideuteron survives the interaction.
Data on these cross sections are sparse, and it is therefore necessary to make approximations based on re-scaling of $\bar p p$ data and use of charge conjugate processes.
This has recently been discussed in detail in Ref.~\cite{Grefe:2015jva}, and we will here adopt their fits to experimental data.

The source term $Q$ is for the case of annihilating dark matter given by
\beq
Q(\vec r, T) = \frac{1}{2}\frac{\rho^2 (\vec r)}{m_{\rm DM}^2} \sum_i \langle \sigma v \rangle_i \frac{dN^i_{\bar d}}{dT},
\eeq
where $\rho(\vec r)$ is the dark matter density, $m_{\rm DM}$ is its mass, and $\langle \sigma v \rangle_i$ is the thermally averaged dark matter annihilation cross section for channel $i$.
For the dark matter halo profile, we chose the Navarro-Frenk-White (NFW)~\cite{Navarro:1995iw} profile,
\beq
\rho(r) = \frac{\rho_0}{(r/r_S)\left[ 1 + \left(r/r_S \right) \right]^2},
\eeq
which has been shown to be in good agreement with the results of N-body halo formation simulations.
For the free parameters in the NFW-profile we use $\rho_0 = 0.26$~GeV/cm$^3$ and $r_S = 20$~kpc.

For the free parameters of the diffusion model, it has been common in the literature to use the three sets of values given in Ref.~\cite{Donato:2003xg}, that yield maximal, 
median and minimal antiproton fluxes from dark matter annihilation, while remaining compatible with the observed B/C ratio.
These parameter sets are labeled `max', `med' and `min' respectively, and their values are listed in Table~\ref{tab:PropModels}.
The `max' and `min' models are often used to estimate the uncertainty band from propagation, but as these are are physically extreme models, the resulting uncertainty band is likely overly conservative.
Indeed, the `min' model has recently been excluded by cosmic ray positron data~\cite{Lavalle:2014kca}.
Propagation uncertainty has been thoroughly discussed in the literature, and is not the focus of this article. We therefore restrict our propagation calculation to the `med' model.

\begin{table}
\begin{tabular}{l c c c c c c c c}
	Model 		&& $L$ in kpc	&& $\delta$	&& $D_0$ in kpc$^2$\,Myr$^{-1}$	&& $V_c$ in km\,s$^{-1}$		\\ 
    \hline
    	max 	&& 15 			&& 0.46 	&& 0.0765						&& 5							\\
    	med		&& 4			&& 0.7		&& 0.0112						&& 12						\\
		min		&& 1			&& 0.85		&& 0.0016						&& 13.5						\\
\end{tabular}
\caption{Propagation parameters for the max, med and min models.}
\label{tab:PropModels}
\end{table}

The diffusion equation~\eqref{eq:Diffusion} can be solved semi-analytically~\cite{Donato:2001ms}, and for annihilating dark matter, the expression for antideuteron flux near Earth is
\beq \label{eq:FluxEarth}
 \Phi_{\bar{d}}(T,\vec{r}_{\odot}) = 
 \frac{v_{\bar{d}}}{4 \pi} \left( \frac{\rho_{0}}{m_{\rm DM}} \right)^2 R(T)
 \frac{ \left<\sigma v \right> }{2} \frac{dN_{\bar{d}}}{dT}
 \,,
\eeq
where
\beq \label{eq:RT}
R(T)=\sum^\infty_{n=1} J_0 \left(\zeta_n \frac{r_{\odot}}{R} \right) \exp\left(-\frac{V_c L}{2K}\right) \frac{y_n(L)}{A_n \sinh(S_nL/2)},
\eeq
\beq \begin{split} \label{eq:yT}
y_n(Z) = &\frac{4}{J_1^2(\zeta_n)R^2}\int^R_0 \D r\ r J_0\left(\frac{\zeta_n r}{R}\right) \int^Z_0 \D z\ \Big\{ \\ 
	 & \exp\left(\frac{V_c(Z-z)}{2D}\right)\sinh\left(\frac{S_n(Z-z)}{2}\right) \left(\frac{\rho(r,z)}{\rho_\odot}\right)^2  \Big\},
\end{split} \eeq
\beq \label{eq:An}
A_n=2h\Gamma_{\rm ann} +V_c +DS_n \coth(S_n L/2),
\eeq
and
\beq \label{eq:Sn}
S_n=\sqrt{\frac{V_c^2}{D^2}+4\frac{\zeta^2_n}{R^2}}.
\eeq

The particle physics of dark matter annihilation and the astrophysics of the propagation are here neatly separated ---
the astrophysics of the propagation is contained within the propagation function $R(T)$, which is completely independent of the particle physics of the annihilation process. This function can thus, independently of the dark matter model in question, be tabulated for a given halo and set of diffusion model parameters,
and later applied to the propagation of antideuterons from any model of (symmetric) dark matter annihilation.

Solar modulations is taken into account using the force field approximation~\cite{Gleeson:1968zza}, shifting the kinetic energy of the particles from $T$ to a kinetic energy near Earth of
$T_\otimes=T-|Ze|\phi_{\rm Fisk}$, where the so-called Fisk potential $\phi_{\rm Fisk}=0.5$\,GV is an effective potential that parametrizes the energy loss from the solar wind.
The corresponding antideuteron flux near Earth is then finally given by
\beq
\Phi_\otimes = \frac{p^2_\otimes}{p^2} \Phi = \frac{2m_{\bar{d}}T_\otimes + T^2_\otimes}{2m_{\bar{d}}T + T^2} \Phi .
\eeq
More realistic modeling, as well as an estimation of uncertainties due to solar modulation has been discussed in detail in Ref.~\cite{Fornengo:2013osa}.


\subsection{Antideuteron flux near Earth}

We generate events for dark matter annihilations into $b \bar b$ and $W^+ W^-$ final states for DM masses of 100~GeV, 500~GeV and 1~TeV using {\tt Herwig++} and {\tt Pythia~8}. 
For the antideuteron formation we use the two sets of best fit values of $p_0$ and $\sigma_0$, based on $pp$ and $e^+e^-$-data, as discussed in Sec.~\ref{sec:fitValues}. 
We assume 100\% branching ratios into the given channels, and use the canonical value of $\langle \sigma v \rangle = 3\e{-26}\ \rm cm^3 s^{-1}$ for the thermally averaged dark matter annihilation cross section.

\begin{figure*}
\includegraphics[width=0.45\textwidth]{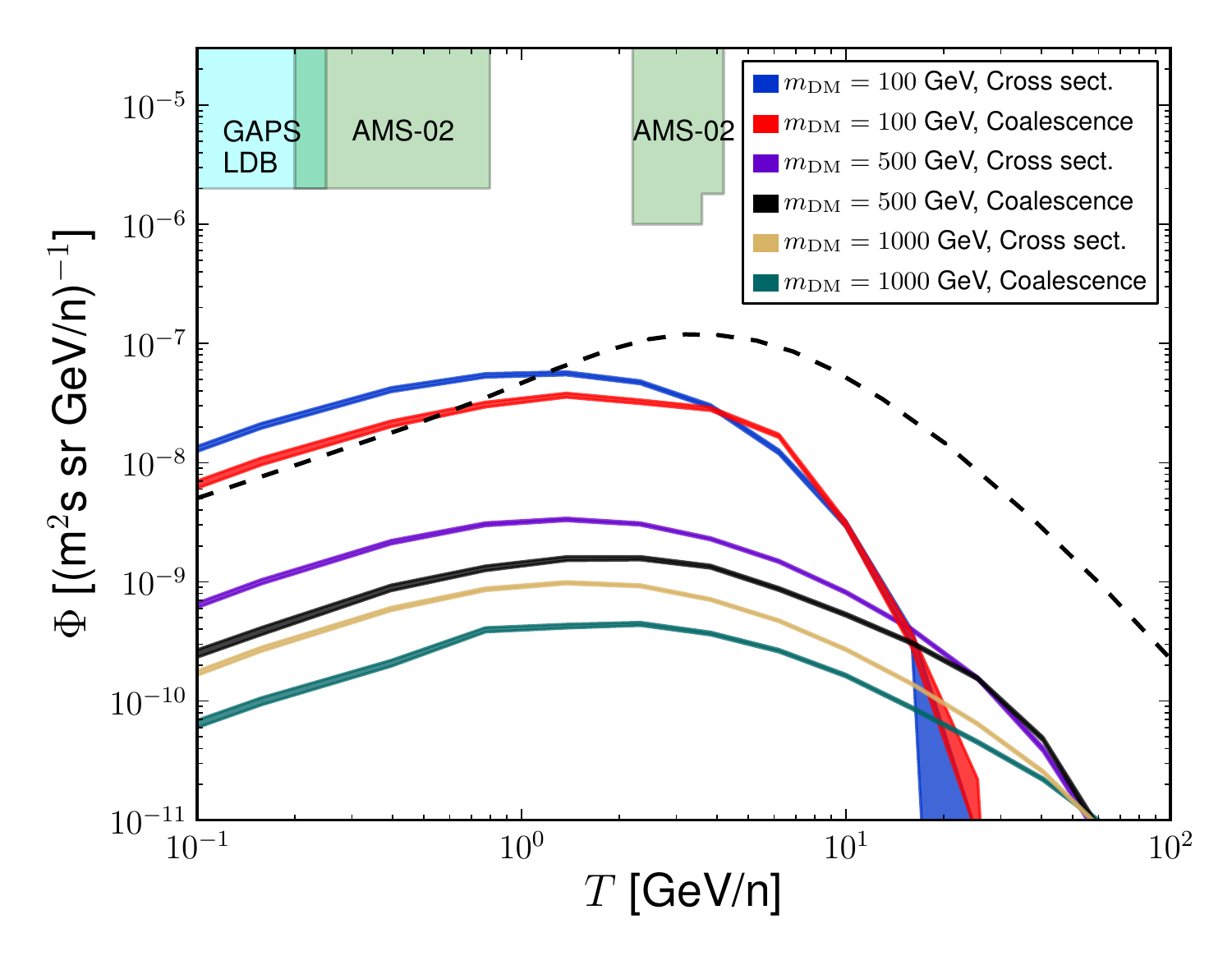}
\includegraphics[width=0.45\textwidth]{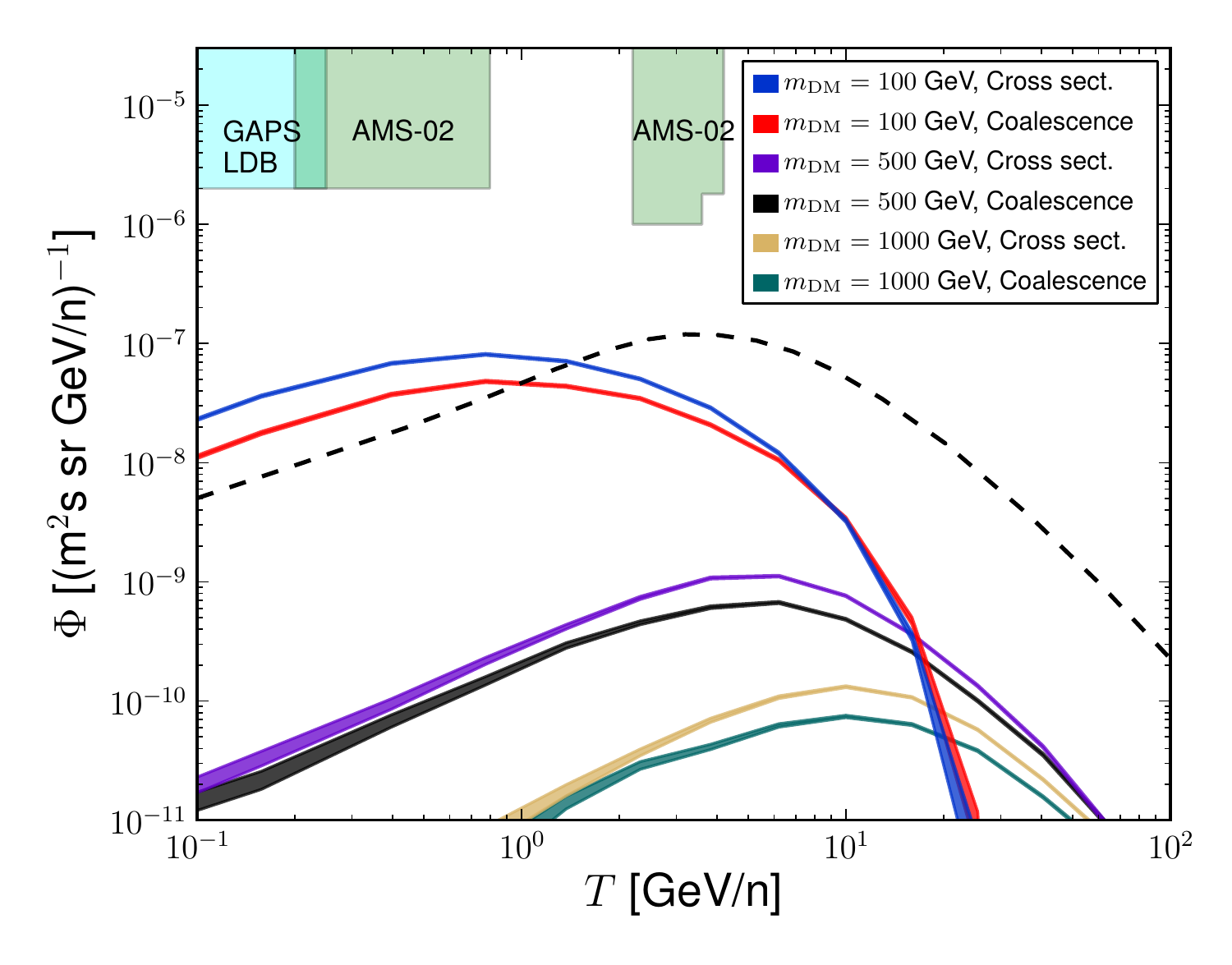}
\caption{Antideuteron spectra at Earth from dark matter annihilation into $b \bar b$ (left) and $W^+W^-$ (right), calculated using {\tt Herwig++} with $p_0 = 96$~MeV in the coalescence model, and $1/\sigma_0 = 0.68\ \rm barn^{-1}$ in the cross section based model. The dashed line shows the expected astrophysical background calculated in Ref.~\cite{Ibarra:2013qt}.}
\label{fig:DM_Herwig_ee}
\end{figure*}

\begin{figure*}
\includegraphics[width=0.45\textwidth]{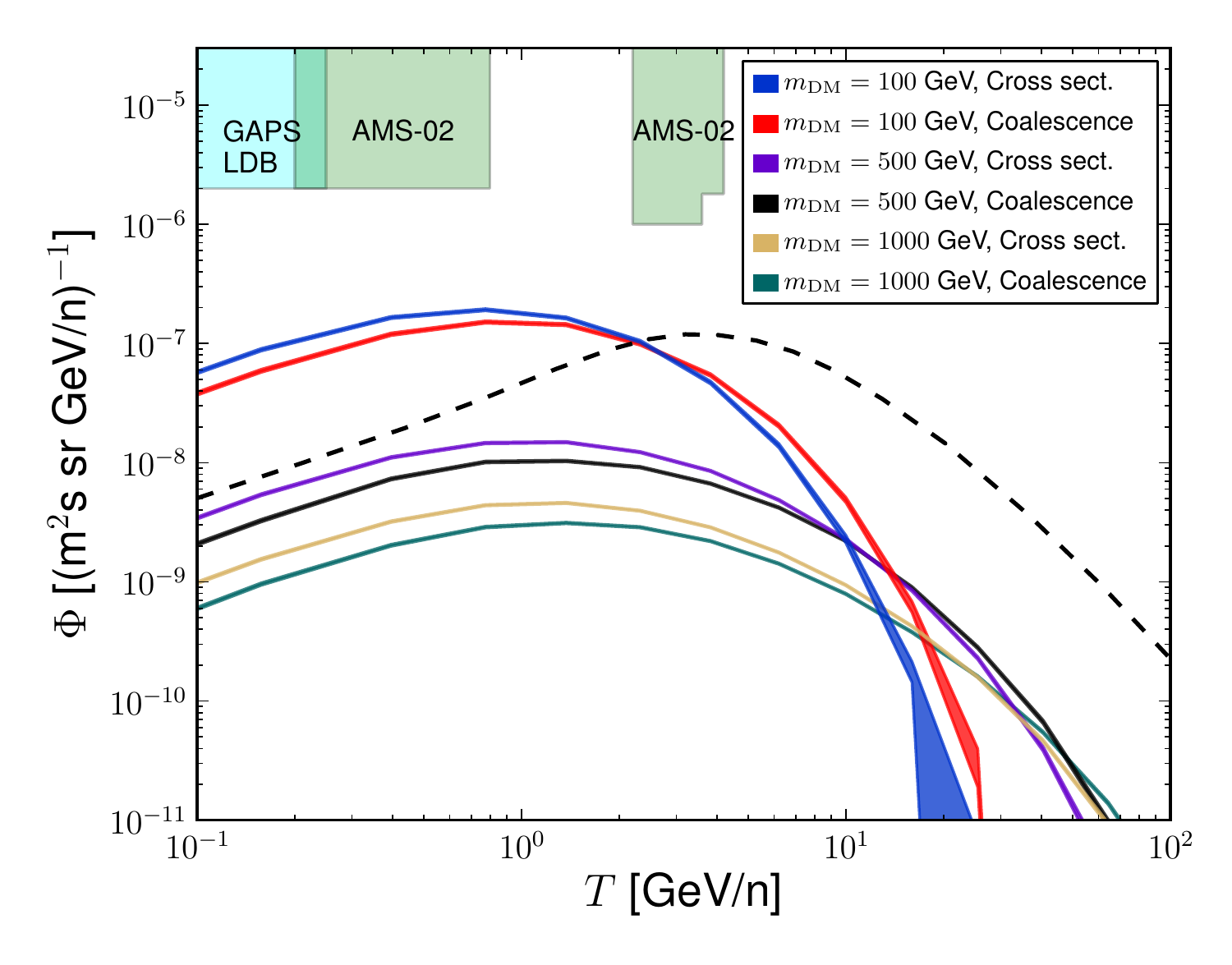}
\includegraphics[width=0.45\textwidth]{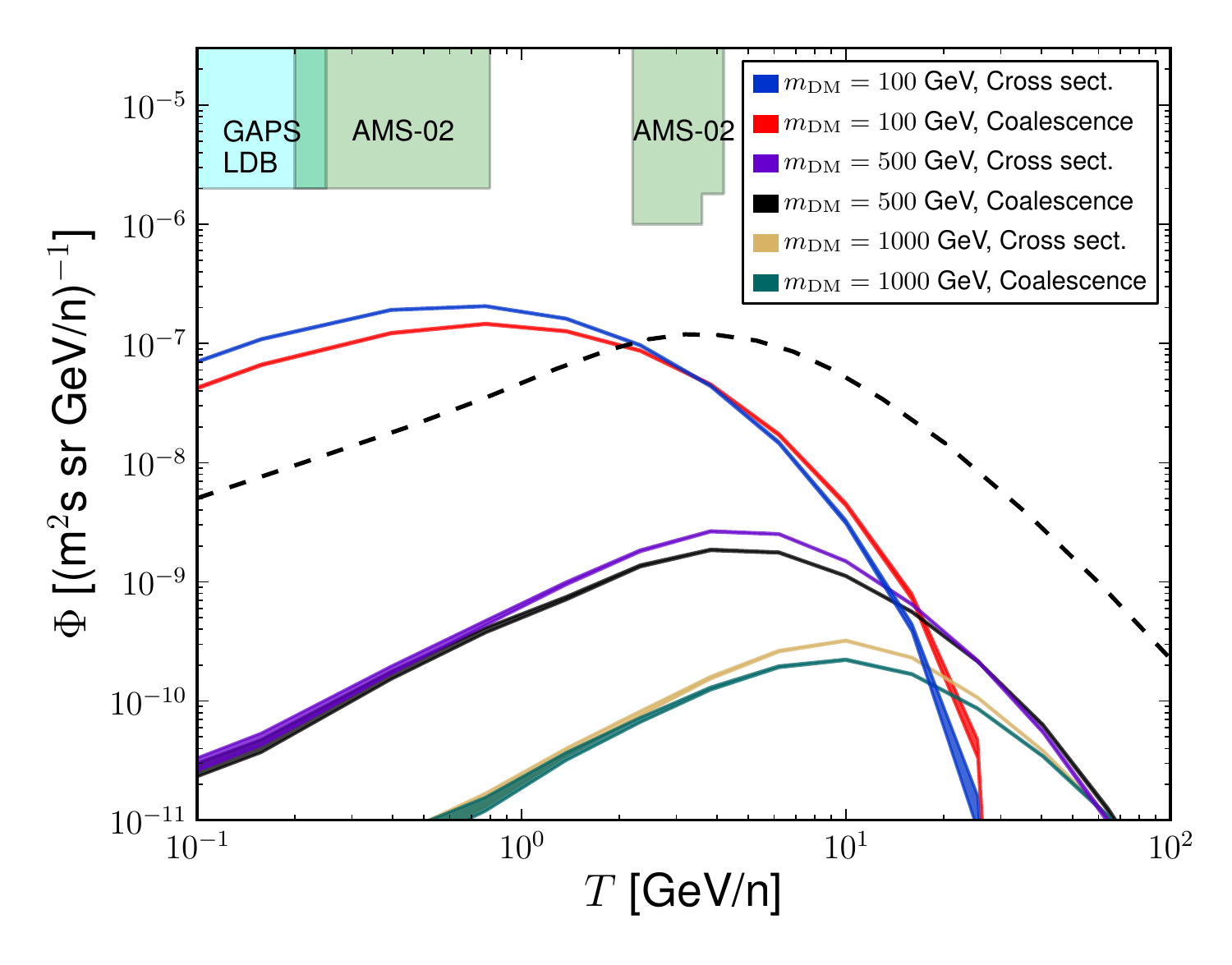}
\caption{Antideuteron spectra at Earth from dark matter annihilation into $b \bar b$ (left) and $W^+W^-$ (right), calculated using {\tt Pythia~8} with $p_0 = 140$~MeV in the coalescence model, and $1/\sigma_0 = 1.18\ \rm barn^{-1}$ in the cross section based model. The dashed line shows the expected astrophysical background calculated in Ref.~\cite{Ibarra:2013qt}.}
\label{fig:DM_Pythia8_ee}
\end{figure*}

In Figs.~\ref{fig:DM_Herwig_ee} and \ref{fig:DM_Pythia8_ee} we, respectively, show the expected antideuteron fluxes after propagation from {\tt Herwig++} and {\tt Pythia 8}.
The figures show the fluxes as a function of the kinetic energy per nucleon of the antideuteron in both the coalescence model and the cross section based model, using the calibration of $\sigma_0$ and $p_0$ against experiments with colorless initial states (BABAR, CLEO, LEP).
The bands indicate the statistical uncertainty in our event generation, and the shaded regions at the top indicate the most recent values for the expected sensitivities of the AMS-02 and GAPS experiments~\cite{Doetinchem}.

\begin{figure*}
\includegraphics[width=0.45\textwidth]{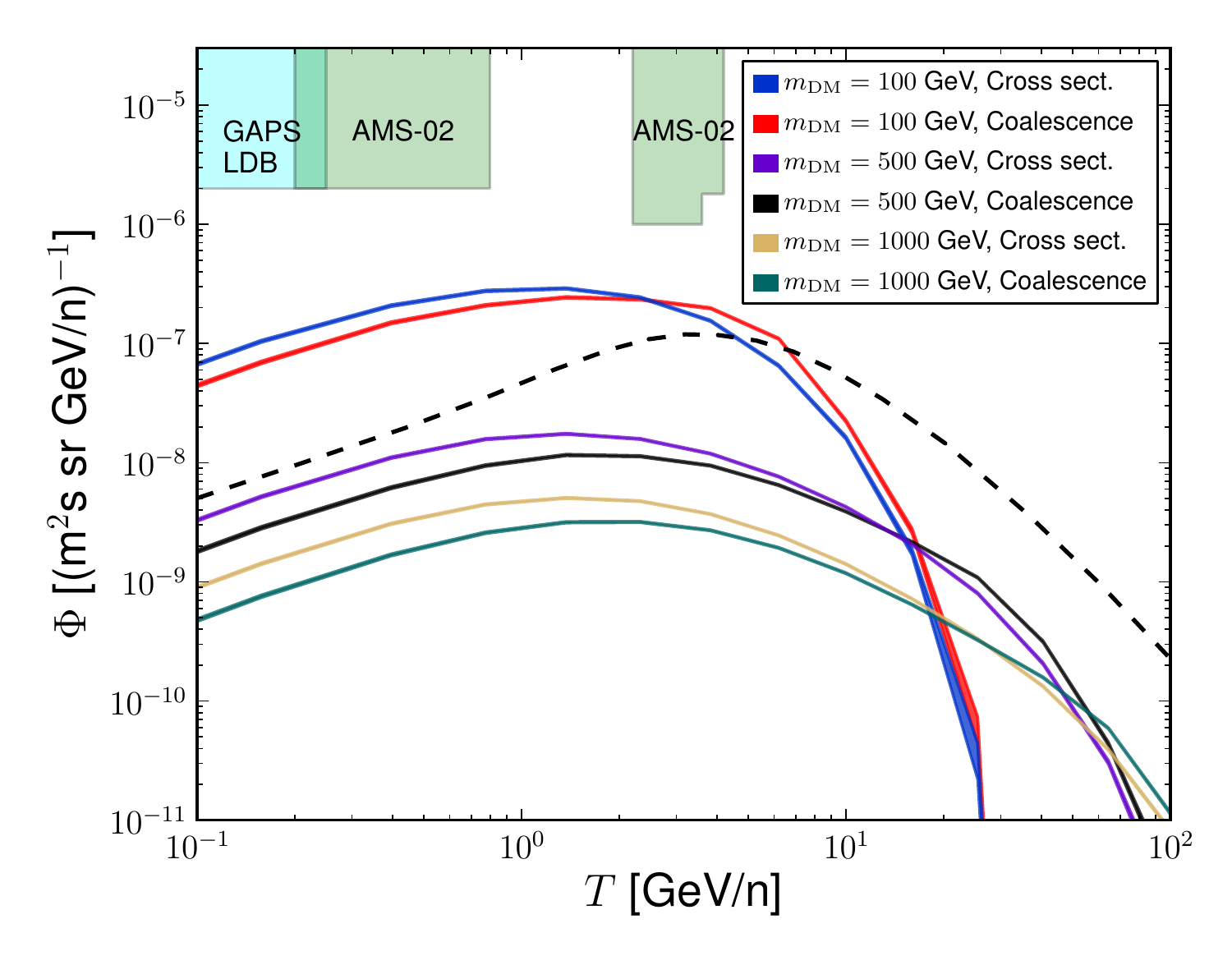}
\includegraphics[width=0.45\textwidth]{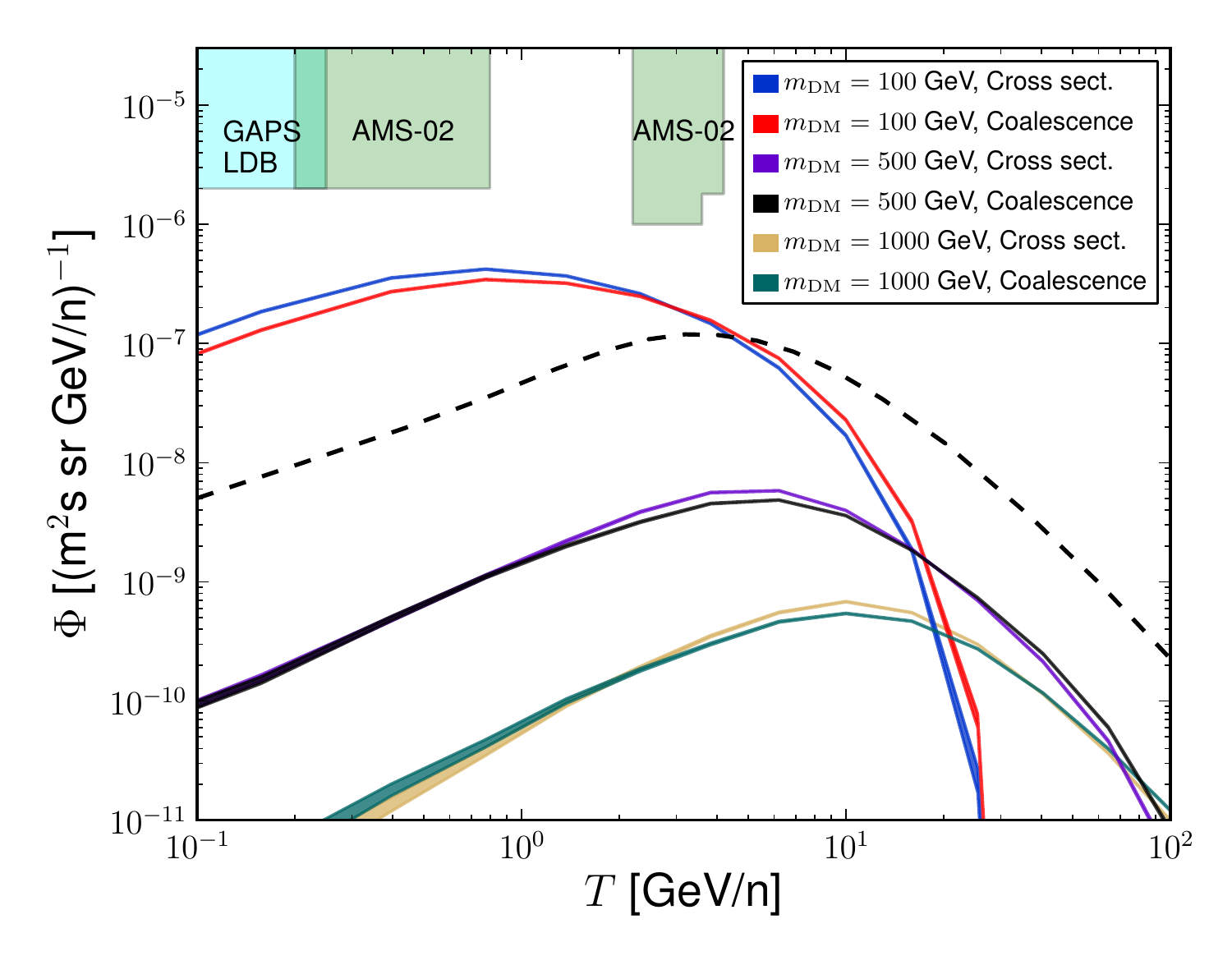}
\caption{Antideuteron spectra at Earth from dark matter annihilation into $b \bar b$ (left) and $W^+W^-$ (right), calculated using {\tt Herwig++} with $p_0 = 187$~MeV in the coalescence model, and $1/\sigma_0 = 3.50\ \rm barn^{-1}$ in the cross section based model. The dashed line shows the expected astrophysical background calculated in Ref.~\cite{Ibarra:2013qt}.}
\label{fig:DM_Herwig_pp}
\end{figure*}

When comparing the predicted fluxes from the two models, one should keep in mind that the relative normalization of the fluxes is not fixed, but determined by the calibration of the free parameters $p_0$ and $\sigma_0$. 
The shapes of the spectra are, however, more or less independent of the calibration, and comparing the shapes thus gives a more reliable picture of the difference between the models.
In particular, the differences between the two models appear larger in {\tt Herwig++} than in {\tt Pythia 8}, but this is largely an effect of the calibration. 
Figure~\ref{fig:DM_Herwig_pp} shows the {\tt Herwig++} result using the calibration against colored initial states (ALICE, ISR),  and we see that the difference between the two models is considerably smaller here due to less of a difference in normalization. In {\tt Pythia 8}, the difference in normalization between the models is similar in the colored and colorless calibrations.
We therefore leave out the plot for the colored calibration in {\tt Pythia~8}.

The differences in the shapes of the spectra between the two models appear to be similar in the two Monte Carlos.
In the $b \bar b$ channel, we see a consistent qualitative difference between the models across all DM masses: the cross section based model predicts a softer antideuteron spectrum, with a more rapid falloff at high energies.
The same can be seen in the 100~GeV dark matter $W^+W^-$ final-state. 
This leads to an enhanced flux in the low energy range relevant for AMS-02 and GAPS, where the background is expected to be small. 
With the values of $p_0$ and $\sigma_0$ used here, the predictions for the flux from the two models typically differ by a factor 1.5--2 in the experimentally relevant energy ranges.

For the higher masses, the situation is less clear for the $W^+W^-$ final-state due to the statistical uncertainty from  limited statistics in the Monte Carlo event generation. 
The two models seem to predict similar slopes at low energies, but the cross section model shows signs of a steeper falloff at high energies.
We see that the cross section model consistently predicts a higher flux at the peak than the coalescence model. 
This leads to a possibly enhanced flux compared to the coalescence model in the multi-GeV kinetic energy region where the AMS-02 experiment has some sensitivity.

\section{Conclusions}
\label{sec:conclusions}
We have proposed a new model for describing the formation of antideuterons in high energy events. Our model is based on the experimentally measured cross sections for nucleon capture processes, and is --- in contrast to the state-of-the-art coalescence model --- capable of describing recent deuteron and antideuteron data from the ALICE experiment at the LHC. 

The physical interpretation of the antideuteron formation process differs significantly between our model and the coalescence model. 
In the coalescence model, antideuteron formation is described by slow nucleon capture, whereas in our model, antideuterons are primarily produced through resonant processes with the delta-resonance, which peaks for COM momentum differences near 1~GeV. 
Moreover, while the coalescence model strictly describes a $\bar p \bar n$ capture process, our model predicts similar antideuteron contributions from $\bar p \bar p$ and $\bar n \bar n$ processes.

We have compared the predictions of our model to the coalescence model for several different experiments, and find our model to give comparable or better descriptions of the data in all experiments; the difference being most significant for the ALICE experiment, where the coalescence model fails to give a satisfactory description.
For the purpose of dark matter indirect detection, we perform fits of the free parameters of the models against two sets of experimental data, divided into experiments with colored and colorless initial states.
We find our model to give consistently better simultaneous fits to the experimental data in both {\tt Herwig++} and {\tt Pythia 8}, and in {\tt Pythia 8}, the fits for our cross section based model give $\chi^2$-values that indicate the model can describe the data successfully.

Comparing the predicted antideuteron fluxes from dark matter annihilation in the two models, with a selection of different dark matter masses and different final states, we find that our model produces softer spectra than the coalescence model, thus giving an enhanced antideuteron flux in the low kinetic energy range relevant for current and planned experiments.

\acknowledgments
We a grateful to Eulogio Serradilla for very helpful discussions on the ALICE analysis, Andy Buckley for insight on the treatment of diffractive and non-diffractive events in event generators, and Philip von Doetinchem for providing estimated sensitivities for the AMS-02 and GAPS experiments.
ARR would like to thank the Cambridge Supersymmetry Working Group, in particular Bryan Webber, for stimulating discussions.
This work was performed on the Abel Cluster, owned by the University of Oslo and the Norwegian metacenter
for High Performance Computing (NOTUR). The computing time was given by NOTUR allocation NN9284K, financed through the Research Council of Norway.


\bibliographystyle{plain}	

\end{document}